\documentclass[lettersize,journal]{IEEEtran}
\usepackage{amsmath,amsfonts}
\usepackage{array}
\ifCLASSOPTIONcompsoc
\usepackage[caption=false,font=footnotesize,labelfont=sf,textfont=sf]{subfig}
\else
  \usepackage[caption=false,font=footnotesize]{subfig}
\fi
\usepackage{textcomp}
\usepackage{stfloats}
\usepackage{url}
\usepackage{verbatim}
\usepackage{graphicx}
\usepackage{cite}
\usepackage{hyperref}
\usepackage{booktabs}
\usepackage{graphicx}         
\usepackage{textcomp}         
\usepackage{xcolor}           
\usepackage{array}            
\usepackage{algorithm}        
\usepackage{algpseudocode}    
\usepackage{diagbox}          
\usepackage{multirow}         
\usepackage{booktabs}         
\usepackage{authblk}          
\usepackage{enumitem}
\begin{document}

\title{Bandwidth-Aware LLM Inference on Heterogeneous Many-Core Supercomputers}
\author{Yao~Lu,
Zhongzhi~Luan, \IEEEmembership{Senior Member,~IEEE},
Gen~Li,
Jiaxing~Qi,
Shiqing~Ma,
Bin~Han,
Shizhe~Shang,
Hailong~Yang, \IEEEmembership{Member,~IEEE},
and~Depei~Qian, \IEEEmembership{Senior Member,~IEEE}
\IEEEcompsocitemizethanks{\IEEEcompsocthanksitem
Yao~Lu, Zhongzhi~Luan, Gen~Li, Jiaxing~Qi, Shiqing~Ma, Bin~Han, Shizhe~Shang, Hailong~Yang, and Depei~Qian are with the Sino-German Joint Software Institute, Beihang University, Beijing 100191, China.\protect\\
E-mail: \{luyuan, luan.zhongzhi\}@buaa.edu.cn.%
\IEEEcompsocthanksitem (Corresponding author: Zhongzhi Luan.)}%
}

\markboth{Preprint}%
{Lu \MakeLowercase{\textit{et al.}}: Bandwidth-Aware LLM Inference on Heterogeneous Many-Core Supercomputers}

\maketitle

\begin{abstract}
Large language model (LLM) inference is limited by high computational cost and memory bandwidth demands, making deployment on heterogeneous many-core processors challenging. Taking the MT-3000 processor used in the Tianhe supercomputer as an example, its limited main memory bandwidth and distributed memory hierarchy exemplify these bottlenecks, making it difficult to directly migrate existing GPU-based inference frameworks. To address this, we propose THInfer, a hardware-aware inference framework that maximizes data locality under bandwidth-constrained conditions through hardware-software co-design and parallel strategy optimization. The framework incorporates three key technologies: (1) a high-performance operator library for the VLIW-SIMD architecture, providing hand-optimized FP16 kernels that achieve up to 70\% of peak performance per cluster; (2) a density-driven computation graph fusion and unified kernel scheduling mechanism, combined with a staged pipelined attention fusion method for co-design; (3) a Prefill–Buffer–Decode (P–B–D) pipeline and bounded buffer management strategy, which supports hybrid parallelism while enabling efficient multi-cluster collaboration and scaling through two-level communication integrating MPI and hthreads. Experiments on the Llama model series show that THInfer improves throughput on the 7B model by 62\%–73\% over DeepSpeed on 2×V100S, and by 67\%–84\% over the A800 GPU. The 13B and 30B models also demonstrate comparable or even better performance. Moreover, THInfer maintains stable performance on the 70B model, whereas typical GPU-based frameworks fail. Overall, THInfer significantly enhances throughput, reduces latency, and improves scalability, providing a feasible technical pathway and system solution for efficient and scalable LLM inference on heterogeneous many-core architectures.
\end{abstract}


\section{Introduction}
In recent years, large language models (LLMs) such as Llama~\cite{touvron2023llama}, Qwen~\cite{bai2023qwen}, and DeepSeek~\cite{guo2025deepseek} have achieved remarkable breakthroughs in natural language processing and multimodal reasoning, owing to their powerful semantic modeling capabilities. As model sizes surge to hundreds of billions of parameters, the computational and memory demands during inference have grown exponentially, imposing extremely high requirements on the computing platform's arithmetic capability, memory bandwidth, and energy efficiency. Existing inference frameworks (e.g., vLLM~\cite{kwon2023efficient}, TensorRT-LLM~\cite{TensorRT-LLM}, and DeepSpeed~\cite{aminabadi2022deepspeed}) improve efficiency through dynamic batching, memory optimization, and quantization. However, their designs heavily rely on high-bandwidth unified memory architectures (such as the 900 GB/s memory bandwidth of NVIDIA V100), making them difficult to adapt to specific high-performance computing systems. Therefore, this paper targets the Tianhe New-Generation supercomputers, aiming to develop an efficient and flexible inference framework for LLMs. The system employs independently developed MT-3000 multi-core digital signal processor (DSP), featuring a heterogeneous many-core architecture with ``16 general-purpose CPUs + 4 acceleration clusters'', leveraging instruction-level parallelism and software-controlled memory to achieve high-efficiency, low-power computation. Although this platform offers abundant heterogeneous computing resources, current applications are largely confined to traditional scientific computing domains such as molecular dynamics~\cite{chen2023large}. Due to differences in architecture and programming models, its resource utilization remains low in deep learning applications, resulting in issues such as load imbalance and hardware idling.
\par
The massive parallel computing capability of this processor provides a hardware foundation for distributed inference of models with tens of billions of parameters. Meanwhile, its unique hardware architecture presents three core challenges for developing an efficient LLM inference framework:
\begin{figure}[!tb]
    \centering
    \includegraphics[width=0.7\columnwidth]{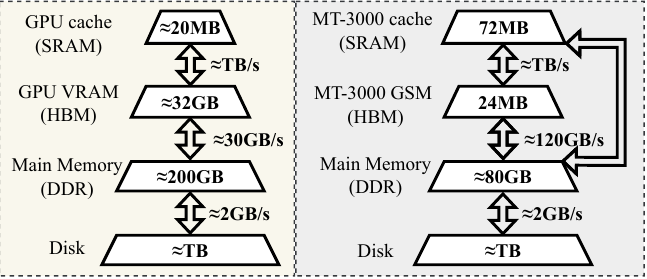} 
    \caption{GPU vs. MT-3000 Memory Hierarchy}
    \label{fig_1}
\end{figure}
\begin{itemize}
   \item \textbf{Unique Hardware Characteristics.} The unique heterogeneous many-core architecture of the MT-3000 leads to a mismatch with traditional computing paradigms. It necessitates fine-grained task partitioning for compute-intensive operators (e.g., GEMM) and memory-intensive operators (e.g., Softmax) across and within accelerator clusters to achieve load balancing and instruction pipeline optimization.
  \item \textbf{Limited Memory Bandwidth.} The actual main memory bandwidth per processor is only about 120GB/s, which is significantly lower than the memory bandwidth of GPUs with comparable computing power (a comparison of their memory hierarchies is shown in Fig.~\ref{fig_1}). Additionally, the limited on-chip storage capacity makes improving data reuse to alleviate bandwidth constraints a key aspect of performance optimization.
  \item \textbf{Scalability bottlenecks.} The orders-of-magnitude gap between the roughly 20 GB of main memory available per cluster and the storage required by models with tens of billions of parameters causes the parallelization strategies of existing distributed inference frameworks to suffer scalability limits due to cross-cluster communication overhead.
\end{itemize}
\par
To address these challenges, we propose \textbf{THInfer}––an LLM inference framework for the Tianhe system. Its core innovation lies in establishing a deep collaborative optimization mechanism between hardware characteristics and model architecture. The framework adopts a co-design approach across three levels: operators, algorithms, and system architecture, addressing the challenges as follows: 
At the \textbf{operator level}, by adapting to hardware features, hand-written assembly kernels based on dataflow analysis are developed for high-frequency operators such as linear and attention layers, increasing single-cluster computational efficiency to 70\% of the theoretical peak. FP16 optimization is employed to reduce memory usage, thereby mitigating hardware mismatch.
At the \textbf{algorithmic level}, by abstracting the model’s computational graph, a “low-density–high-density–low-density” operator fusion strategy is adopted, embedding low-density operators like RoPE into high-density kernels. Furthermore, a phased MT Attention mechanism is proposed to align with the storage and broadcasting characteristics of AM/SM/GSM, reducing I/O access pressure. Unified kernels are introduced to minimize launch overhead, effectively addressing limited memory bandwidth and insufficient computational efficiency. 
At the \textbf{system architecture level}, a three-stage synchronized pipeline based on Prefill–Decode separated scheduling, namely P–B–D (Prefill–Buffer–Decode), is constructed. This integrates a selective batching mechanism and combines KV cache asynchronous migration with computation overlap to prevent request accumulation and out-of-memory (OOM) errors. A constraint-based hybrid parallel strategy is employed during both Prefill and Decode stages: intra-cluster synchronization is achieved via hthread, while inter-cluster communication relies on MPI, forming a two-level parallel mode that collaboratively enables efficient aggregation of cross-device memory bandwidth, thereby overcoming scalability bottlenecks. 
\par
Experimental results demonstrate that the framework achieves near-linear throughput scaling on commonly used LLMs such as \textbf{Llama}, while significantly reducing end-to-end inference latency. Under peak performance-aligned experimental settings, the end-to-end throughput of this framework is overall comparable to or better than the \textbf{V100/A800 baselines based on DeepSpeed}, providing an efficient and feasible solution for industrial-scale LLM inference on the Tianhe New-Generation supercomputers.
\par
The main contributions of this paper can be summarized as follows:
\begin{itemize}
    \item \textbf{High-Performance Operator Library for MT-3000.} Developing dataflow analysis-based high-performance operators and FP16 kernels tailored for the MT-3000 architecture, co-optimizing computational efficiency and memory usage;
    \item \textbf{Density-Aware Deep Computation Graph Fusion.} Proposing a density-driven fusion strategy to embed low-density operators into high-density kernels, and designing an MT Attention mechanism adapted to memory hierarchy for reduced I/O and scheduling overhead;
    \item \textbf{Adaptive Parallel Scheduling Mechanism.} Constructing a P--B--D three-stage pipeline at micro-batch granularity to safely decouple prefill and decode phases. Combined with hybrid parallelism and communication optimization, it supports efficient scaling for models with tens of billions of parameters;
    \item \textbf{Systematic Evaluation Framework.} Establishing a multi-dimensional performance evaluation system that validates the superiority and robustness of the proposed approach in terms of latency, throughput, and resource utilization on the Llama series of models.
\end{itemize}
\section{Background and motivation}
\subsection{Tianhe New-Generation Supercomputer}
The Tianhe new-generation supercomputer is built on the independently developed MT-3000 DSP, employing an innovative CPU-DSP hybrid architecture for hardware design~\cite{lu2022mt}. This processor achieves an energy efficiency ratio of 45.4 GFLOPS/W in double-precision mode, representing a 62\% improvement compared to the NVIDIA V100~\cite{khalilov2021performance}. Its prototype system delivers a peak performance of 12 Petaflops with a Linpack efficiency of 80\%, demonstrating significant large-scale scalability.
\par
\begin{figure*}[!tb]
  \centering
  \subfloat[Overview of MT-3000 Architecture]{%
    \includegraphics[height=3.5cm]{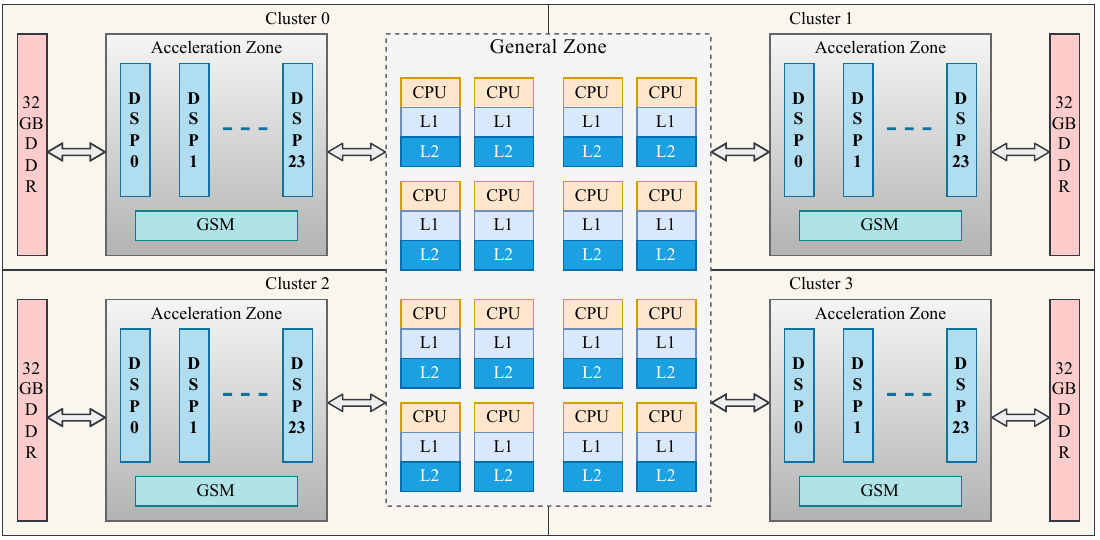}\label{fig:1a}}
  \hspace{0.01\linewidth}
  \subfloat[DSP Internal Structure]{%
    \includegraphics[height=3.5cm]{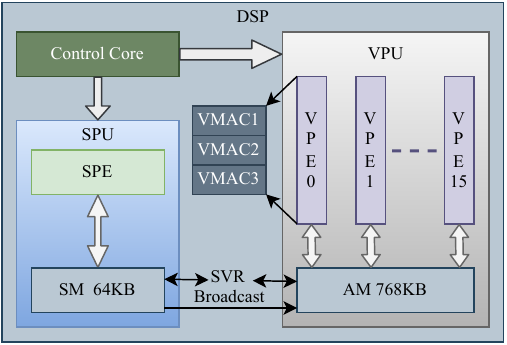}\label{fig:1b}}
  \caption{Illustration of the MT-3000 system.}\label{fig:mt3000}
\end{figure*}

The MT-3000 processor is divided into a general-purpose region and an accelerator region: the general-purpose area includes 16 CPUs equipped with two levels of private caches; the acceleration area consists of 4 autonomous clusters, each containing 24 control cores and 384 accelerator cores, with communication achieved through a hierarchical interconnection network, as shown in Fig.~\ref{fig:1a}. Each DSP core is composed of 1 control core and 16 accelerator cores, supporting 1024-bit SIMD instructions and VLIW architecture, integrating a Scalar Processing Unit (SPU), Vector Processing Unit (VPU), and equipped with dedicated storage resources (64KB scalar memory (SM) and 768KB vector memory (AM)). The structure of a single DSP is shown in Fig.~\ref{fig:1b}. 

\par
The MT-3000 adopts a three-tier memory hierarchy: on-chip AM/SM cache directly interfaces with registers, the 6MB GSM within the cluster supports cross-core data sharing, while off-chip DDR memory enables efficient data transfer via DMA, with bandwidth ranked as AM/SM > GSM > DDR. DMA supports asynchronous transfer and broadcast mechanisms. The communication system adopts a hybrid programming model: within clusters, dynamic multi-core task scheduling is achieved through the hthreads heterogeneous threading library, and inter-cluster collaboration is based on the MPI protocol~\cite{walker1996mpi} for data synchronization.
\subsection{Classical LLMs}
Since the introduction of the Transformer architecture~\cite{vaswani2017attention}, its powerful contextual modeling and cross-task generalization capabilities have driven a paradigm shift in artificial intelligence. In natural language processing (NLP), BERT~\cite{koroteev2021bert}, based on bidirectional masked language modeling, broke through semantic representation bottlenecks with 340 million parameters. The GPT series started with GPT-2's~\cite{radford2019language} 1.5 billion parameters, demonstrating that large-scale autoregressive models could significantly improve text generation quality. GPT-3~\cite{brown2020language} reached 175 billion parameters, exhibiting emergent abilities and few-shot learning for the first time. The Llama~\cite{touvron2023llama, touvron2023llama2, dubey2024llama} series introduced Grouped Query Attention (GQA), achieving leading multi-task performance within a parameter range of 7 billion to 405 billion. DeepSeek-V3~\cite{liu2024deepseek} employs a Mixture of Experts (MoE) architecture, dynamically activating 37 billion parameters with an effective capacity of 671 billion, making it suitable for complex reasoning tasks such as mathematics and programming. The Qwen~\cite{bai2023qwen, team2024qwen2} series supports multimodal tasks through cross-modal alignment, including visual question answering and document analysis. Research shows a significant positive correlation between model scale and performance—scaling parameters from hundreds of millions to hundreds of billions not only enhances generation quality but also catalyzes emergent capabilities.
\subsection{LLM Inference Optimization Techniques}
To enhance the inference efficiency of LLMs, various optimization techniques have been proposed. In model compression and quantization, methods such as GPTQ~\cite{frantar2022gptq} and AWQ~\cite{lin2024awq} achieve aggressive 3–4 bit quantization, reducing resource requirements; SparseGPT~\cite{frantar2023sparsegpt} combines sparsification and quantization to simultaneously decrease computational and memory overhead. Pruning and distillation techniques (e.g., LLM-Pruner~\cite{ma2023llm}, MiniLLM~\cite{gu2023minillm}) compress model scale through structured pruning and knowledge distillation. For attention mechanism optimization, Flash Attention~\cite{dao2022flashattention} reduces computational complexity via tiling and I/O optimization; models like Longformer~\cite{beltagy2020longformer}, Linformer~\cite{wang2020linformer}, and Reformer~\cite{kitaev2020reformer} leverage sparsification, low-rank approximation, or hashing techniques to lower complexity to linear or near-linear. To address the KV cache memory bottleneck, PagedAttention~\cite{kwon2023efficient} improves throughput via paged management; GQA~\cite{ainslie2023gqa} reduces the number of key-value heads; SCISSORHANDS~\cite{liu2023scissorhands} prunes redundant tokens; and StreamingLLM~\cite{xiao2023efficient} compresses memory using a combination of window attention and sink token mechanisms. At the system level, scheduling strategies such as ORCA~\cite{yu2022orca} enhance throughput and concurrency through iteration-level scheduling and selective batching. Together, these techniques advance LLMs toward greater efficiency and practicality, laying the foundation for edge computing and real-time applications.
\subsection{LLM Inference Frameworks}
Transformer inference faces significant computational and memory bottlenecks, for which various frameworks have proposed efficient solutions. \textbf{vLLM} mitigates memory fragmentation through \textbf{PagedAttention}, improving throughput by 2--4$\times$~\cite{kwon2023efficient}; \textbf{TensorRT-LLM} leverages TensorRT for deep optimization, supporting dynamic batching and quantization for low-latency inference on NVIDIA GPUs~\cite{TensorRT-LLM}; \textbf{TGI} employs continuous batching and tensor parallelism as a high-performance inference backbone in the Hugging Face ecosystem~\cite{huggingface_tgi}; \textbf{DeepSpeed Inference} optimizes multi-GPU and heterogeneous computing to enable real-time inference for trillion-parameter models~\cite{aminabadi2022deepspeed}; \textbf{FlexGen} enhances heterogeneous resource scheduling with 4-bit quantization for high-batch offline tasks~\cite{sheng2023flexgen}; and \textbf{DistServe} decouples \textbf{Prefill} and \textbf{Decode} stages through bandwidth-aware placement and pipeline scheduling, combining continuous batching with cross-request KV reuse to improve goodput under \textbf{TTFT/TPOT} constraints~\cite{zhong2024distserve}. These advancements collectively address key challenges in scalable and efficient LLM deployment.
\section{Methods}
\subsection{THInfer Framework}
\begin{figure*}[!tb]
    \centering
    \includegraphics[width=0.9\textwidth]{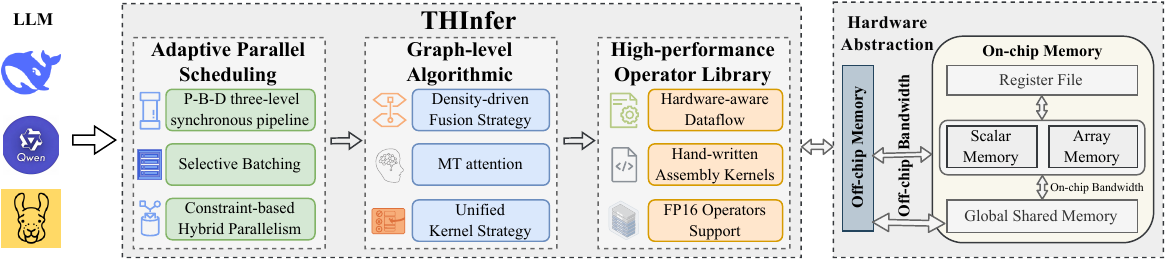} 
    \caption{THInfer: Co-Designing Operators, Graph-Level Algorithms, and System-Level Adaptive Parallelism for MT-3000 LLM Inference}
    \label{fig_3}
\end{figure*}
The THInfer framework adopts a systematic co-optimization approach, constructing a multi-level LLM inference system with deep hardware integration. Its overall architecture is illustrated in Fig.~\ref{fig_3}.
\par
\textbf{Operator-Level Optimization:} As the foundation of system performance, this layer focuses on unlocking the extreme performance of key operators (e.g., GEMM). Through hand-written assembly kernels, fine-grained dataflow design, and FP16 precision support, it fully leverages the computational potential of the single-cluster VLIW-SIMD architecture, effectively translating theoretical throughput into actual performance.
\par
\textbf{Algorithm-Level Optimization:} Building upon the operator layer, we break through the traditional operator-centric scheduling paradigm. By employing a density-driven fusion strategy, multiple fine-grained operators are fused into unified macro-operators. Moreover, we innovatively propose MT Attention to replace Flash Attention, significantly reducing redundant memory accesses and kernel launch overhead while improving on-chip memory utilization.
\par
\textbf{System-Level Adaptive Parallel Optimization:} At the highest level, we restructure the entire inference pipeline from a system-wide perspective. A three-stage P-B-D synchronous pipeline is designed, integrating selective batching and hybrid parallel strategies. This ensures safe decoupling of the prefill and decoding phases while enabling efficient coordination and scaling of multi-cluster resources through techniques such as asynchronous KV-Cache migration and computation-communication overlap.
\subsection{High-performance Operator Design}
\begin{figure}[!tb]
    \centering
    \includegraphics[width=\columnwidth]{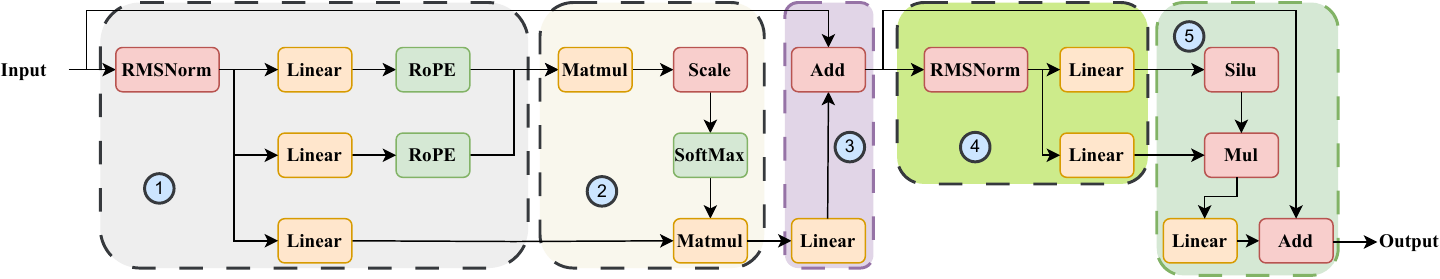}
    \caption{Computational Flowchart of LLM inference}
    \label{fig_4}
\end{figure}
Based on the abstract operator graph of LLMs (Fig.~\ref{fig_4}), operators are categorized into high-density (e.g., GEMM) and low-density (e.g., Softmax/RoPE) types according to their arithmetic intensity~\cite{sze2017efficient}. The Roofline model indicates that high-density operators are often compute-bound, while low-density operators are mostly I/O-bound~\cite{williams2009roofline}. Accordingly, double buffering is used for low-density operators to overlap computation and data transfer, while dataflow and kernel-level optimizations are applied to high-density operators to improve computational utilization.
\subsubsection{GEMM Operator Analysis} 
Consider matrix multiplication \( Y = X \times W \), where \( X \in \mathbb{R}^{M \times K} \), \( W \in \mathbb{R}^{K \times N} \), \( Y \in \mathbb{R}^{M \times N} \). Combining hardware parameters (memory bandwidth \( B = 30 \) GB/s, FP32 peak multiply-accumulate performance \( \mathrm{FP32_{peak}} = 4.05 \) TFLOPS), the following analysis can be performed.
For FP32, the computational intensity is
\begin{equation}
\label{eq:intensity-fma}
I = \frac{\mathrm{FLOPs}}{D_{\text{bytes}}} = \frac{M K N}{4 (M K + K N + M N)}.
\end{equation}
\par
The roofline intensity threshold is
\begin{equation}
\label{eq:intensity-threshold}
I_{\text{threshold}} = \frac{FP32_{\text{peak}}}{B} = \frac{4.05 \times 10^{12}}{30 \times 10^9} = 135\text{FLOPs/Byte}.
\end{equation}
\par
The attainable performance is
\begin{equation}
\label{eq:pmax}
P_{\max} = \min\left(FP32_{\text{peak}}, B \cdot I \right).
\end{equation}
\par
Due to the unique on-chip memory structure of the MT-3000 processor, this paper employs the outer product method to optimize the GEMM operation~\cite{yu2024optimizing}. Under the capacity constraints of the SM and AM in FP32 precision, the following conditions must be satisfied.
\begin{equation}
4B \times M \times K \leq 64 \text{ KB}.
\end{equation}
\begin{equation}
4B \times K \times N + 4B \times M \times N \leq 768 \text{ KB}.
\end{equation}
\par
Under these constraints, we obtain an optimal set of configuration parameters: \( M = 128 \), \( K = 128 \), \( N = 768 \). The computational intensity \( I \) at this point is \( 192/13 \), which is significantly lower than the threshold \( I_{\mathrm{threshold}} \), indicating that the performance is primarily limited by memory bandwidth.
\subsubsection{Dataflow Design and Memory Access Optimization}
\begin{figure}[!tb]
    \centering
    \includegraphics[width=0.6\columnwidth]{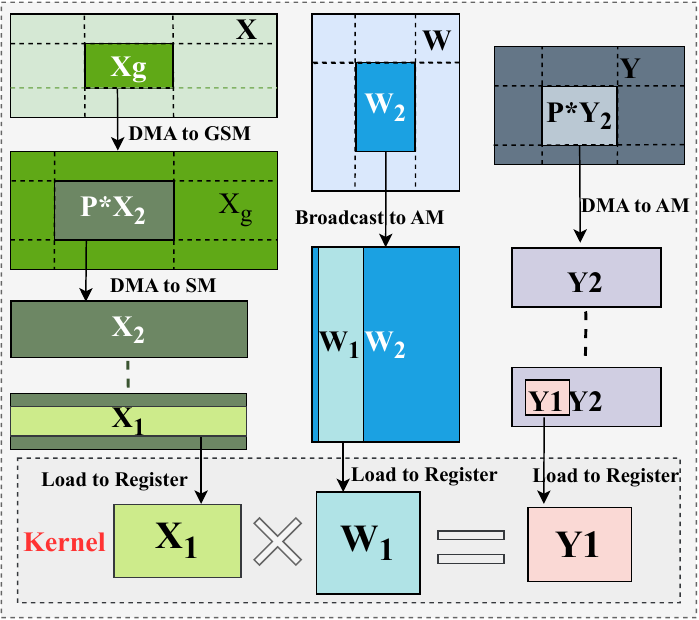}
    \caption{Data Flow Graph for GEMM Operator Computation: Three-level dynamic tiling (DDR→GSM→AM) with triple buffering for Linear operators, optimizing dataflow via W broadcasting and static output Y to reduce DMA transfers.}
    \label{fig_5}
\end{figure}

For the GEMM operator, when $M$ is large, we partition the input matrix $\mathbf{X} \in \mathbb{R}^{M \times K}$ along the $M$ dimension. The entire process consists of three levels: first, $\mathbf{X}$ is divided into blocks $\mathbf{X}_g[M_g, K_g]$ and cached into GSM leveraging its high-bandwidth characteristics; subsequently, it is decomposed into $P$ pieces of $\mathbf{X}_2[M_2, K_2]$ loaded into SM, and $\mathbf{W}_2[K_2, N_2]$ is broadcast to the AM of each DSP; finally, at the kernel level, $\mathbf{X}_2$ and $\mathbf{W}_2$ are further decomposed into $\mathbf{X}_1[M_1, K_1]$ and $\mathbf{W}_1[K_1, N_1]$ to compute the submatrix $\mathbf{Y}_1[M_1, N_1]$. This process is described in detail in the next section. The workflow is illustrated in Fig.~\ref{fig_5}. To reduce the amount of DMA data transfer, a matrix $\mathbf{W}$ broadcasting method is employed for acceleration, with the output matrix $\mathbf{Y}$ set as a stationary target.
\par
\begin{figure*}[!tb]
    \centering
    \includegraphics[width=0.8\textwidth]{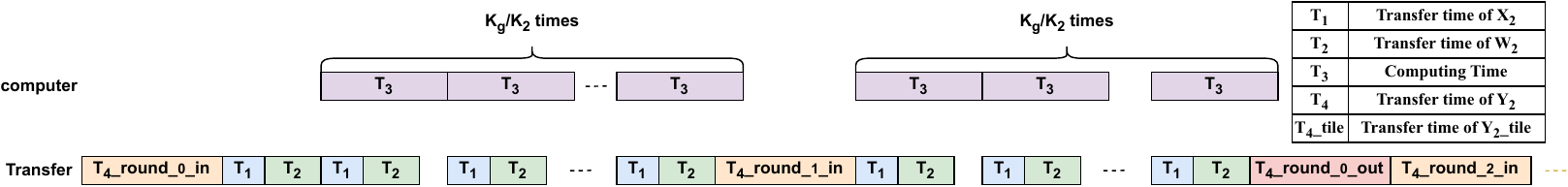}
    \caption{Before Tiling: Unbalanced transmission load of \(Y_2\) input and output}
    \label{fig5a}
\end{figure*}
\begin{figure*}[!tb]
    \centering
    \includegraphics[width=0.6\textwidth]{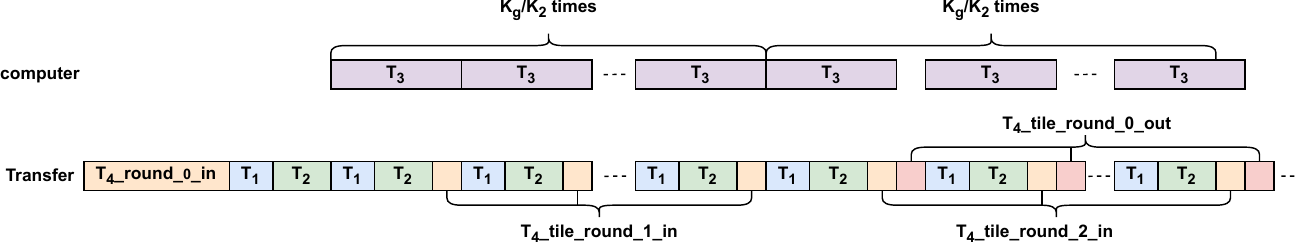}
    \caption{After Tiling: \(Y_2\) input and output are tiled to balance the transmission load}
    \label{fig5b}
\end{figure*}

To maximize parallelism between computation and data transfer, this paper designs a triple-buffering mechanism that decouples the input, output, and computation processes of $Y_2$, thereby achieving parallelization of computation and data I/O. However, since $Y_2$ is generated through multiple iterations of $X_2$ and $W_2$, it tends to cause uneven transfer loads and induce tail effects (Fig.~\ref{fig5a}). To address this, a dynamic tiling strategy is introduced to further decompose $Y_2$ into finer-grained $Y_2$\_tile and embed its transfer operations into the inner computation loop of each $X_2$ and $W_2$ iteration, as illustrated in Fig.~\ref{fig5b}.

\par
Based on the storage capacity constraints of the MT-3000, the optimization parameters need to meet the capacity constraints of SM, AM, and GSM.
\begin{equation}
\label{formula3}
2 \times 4B \times M_2 \times K_2 \leq 64 \text{KB},
\end{equation}
\begin{equation}
\label{formula4}
2 \times 4B \times K_2 \times N_2 + 3 \times 4B \times M_2 \times N_2 + 4B \times N \leq 768 \text{KB},
\end{equation}
\begin{equation}
\label{formula5}
4B \times K_g \times M_g \leq 6 \text{MB}.
\end{equation}
\par
To enhance computational efficiency, the kernel size is set to \( M_{1} = 6 \) and \( N_{1} = 128 \), with \( N_{1} = 128 \) chosen to better align with the Multi-Head Attention (MHA) mechanism. The optimal configuration derived from parameter exploration is given by \( M_g = 720 \), \( K_g = 2,\!048 \), \( M_2 = 30 \), with \( K_2 \) and \( N_2 \) both set to 256.
\par
When the dimension $M$ is small, we adopt a strategy of broadcasting the input matrix $\mathbf{X}$ and perform parallel computation along the $N$ dimension to achieve load balancing among multiple DSPs.
\subsubsection{VLIW-SIMD instruction-level optimization}
To fully exploit the computational potential of the MT-3000 processor and address the suboptimal performance of the compiler-generated outer product kernel, this section proposes an optimization method for matrix multiplication based on hand-written assembly. Each accelerator core of the MT-3000 is equipped with three parallel floating-point multiply-accumulate units (VMACs). By offloading computational tasks to 16 accelerator cores via SIMD instructions, each core can efficiently handle matrix operations of size \(6 \times K_2 \times 8\). The detailed optimization strategies are as follows:
\par
A two-row two-column outer product computation pattern is employed. For the left matrix \(X_1\), scalar loading, bit packing, and vector broadcasting are achieved through an instruction chain consisting of \(\text{sldh} \to \text{sbale2} \to \text{svbcast}\), while the right matrix \(W_1\) is directly loaded into vector registers using the \(\text{vldw}\) instruction. Vectorized floating-point multiply-accumulate operations are then executed via the \(\text{vfmulas32}\) instruction, with the instruction pipeline scheduling outlined in Table~\ref{tab:pipeline_schedule}.
\begin{table*}[h!]
\caption{Instruction Pipeline Scheduling for Matrix Multiplication Kernel}
\label{tab:pipeline_schedule}
\centering
\setlength{\tabcolsep}{4pt}
\begin{tabular}{@{}lllll@{}}
\toprule
\textbf{VMAC} & \textbf{SMAC} & \textbf{SLDST} & \textbf{VLDST} & \textbf{SIEU} \\ 
\midrule
vfmulas32 $X_1[0,1,2][0]$, $W_1[0][0]$ & -- & -- & vldw $W_1[1][2,3]$ & -- \\ 
vfmulas32 $X_1[0,1,2][0]$, $W_1[0][1]$ & -- & -- & -- & sbale $X_1[0][1]$ \\ 
vfmulas32 $X_1[0,1,2][0]$, $W_1[0][2]$ & svbcast $X_1[0][1]$ & sldh $X_{1_{\text{next}}}[0][0]$ & -- & sbale $X_1[1][1]$ \\ 
vfmulas32 $X_1[0,1,2][0]$, $W_1[0][3]$ & svbcast $X_1[1][1]$ & sldh $X_{1_{\text{next}}}[1][0]$ & -- & sbale $X_1[2][1]$ \\ 
vfmulas32 $X_1[3,4,5][0]$, $W_1[0][0]$ & svbcast $X_1[2][1]$ & sldh $X_{1_{\text{next}}}[2][0]$ & -- & sbale $X_1[3][1]$ \\ 
vfmulas32 $X_1[3,4,5][0]$, $W_1[0][1]$ & svbcast $X_1[3][1]$ & sldh $X_{1_{\text{next}}}[3][0]$ & -- & sbale $X_1[4][1]$ \\ 
vfmulas32 $X_1[3,4,5][0]$, $W_1[0][2]$ & svbcast $X_1[4][1]$ & sldh $X_{1_{\text{next}}}[4][0]$ & -- & sbale $X_1[5][1]$ \\ 
vfmulas32 $X_1[3,4,5][0]$, $W_1[0][3]$ & svbcast $X_1[5][1]$ & sldh $X_{1_{\text{next}}}[5][0]$ & vldw $W_{1_{\text{next}}}[0][0,1]$ & -- \\ 
vfmulas32 $X_1[0,1,2][1]$, $W_1[1][0]$ & -- & -- & vldw $W_{1_{\text{next}}}[0][2,3]$ & -- \\ 
vfmulas32 $X_1[0,1,2][1]$, $W_1[1][1]$ & -- & -- & -- & sbale $X_{1_{\text{next}}}[0][0]$ \\ 
vfmulas32 $X_1[0,1,2][1]$, $W_1[1][2]$ & svbcast $X_{1_{\text{next}}}[0][0]$ & sldh $X_{1_{\text{next}}}[0][1]$ & -- & sbale $X_{1_{\text{next}}}[1][0]$ \\ 
vfmulas32 $X_1[0,1,2][1]$, $W_1[1][3]$ & svbcast $X_{1_{\text{next}}}[1][0]$ & sldh $X_{1_{\text{next}}}[1][1]$ & -- & sbale $X_{1_{\text{next}}}[2][0]$ \\ 
vfmulas32 $X_1[3,4,5][1]$, $W_1[1][0]$ & svbcast $X_{1_{\text{next}}}[2][0]$ & sldh $X_{1_{\text{next}}}[2][1]$ & -- & sbale $X_{1_{\text{next}}}[3][0]$ \\ 
vfmulas32 $X_1[3,4,5][1]$, $W_1[1][1]$ & svbcast $X_{1_{\text{next}}}[3][0]$ & sldh $X_{1_{\text{next}}}[3][1]$ & -- & sbale $X_{1_{\text{next}}}[4][0]$ \\ 
vfmulas32 $X_1[3,4,5][1]$, $W_1[1][2]$ & svbcast $X_{1_{\text{next}}}[4][0]$ & sldh $X_{1_{\text{next}}}[4][1]$ & -- & sbale $X_{1_{\text{next}}}[5][0]$ \\ 
vfmulas32 $X_1[3,4,5][1]$, $W_1[1][3]$ & svbcast $X_{1_{\text{next}}}[5][0]$ & sldh $X_{1_{\text{next}}}[5][1]$ & vldw $W_{1_{\text{next}}}[1][0,1]$ & -- \\ 
\bottomrule
\end{tabular}
\vspace{0.5em}
\parbox{\textwidth}{\scriptsize\itshape Note: The left matrix $X_1$ is loaded through \texttt{sldh→sbale2→svbcast} chain for scalar loading, bit-packing, and vector broadcasting. The right matrix $W_1$ is directly loaded via \texttt{vldw}.}
\end{table*}

Here, \textbf{VMAC} denotes floating-point multiply-accumulate units, \textbf{SMAC} represents vector multiply-accumulate units, \textbf{SLDST} and \textbf{VLDST} signify scalar and vector load/store units respectively, and \textbf{SIEU} stands for scalar integer execution units. Notably, in this work, every two 32-bit data elements of $W_1$ are combined into 64-bit entities, such that each row of the right matrix processed by each accelerator core contains four double-precision values. The functionality and cycle counts of each instruction are summarized in Table~\ref{tab:instr_cycles}.

\begin{table}[h!]
\caption{Instruction Characteristics}
\label{tab:instr_cycles}
\centering
\renewcommand{\arraystretch}{1.1}
\setlength{\tabcolsep}{3pt}
\resizebox{\columnwidth}{!}{%
\begin{tabular}{@{}lcp{5.5cm}@{}}
\toprule
\textbf{Instruction} & \textbf{Cycle Count} & \textbf{Description} \\
\midrule
\texttt{vfmulas32} & 6 & FP32-precision vector floating-point multiply-accumulate \\
\texttt{svbcast}   & 4 & Broadcast scalar register value to vector register \\
\texttt{sldh}      & 7 & Load single FP32 value into scalar register \\
\texttt{sbale2}    & 1 & Pack lower 32 bits of two scalar registers into one scalar register \\
\texttt{vldw}      & 9 & Load 16$\times$64-bit values into vector register \\
\bottomrule
\end{tabular}%
}
\end{table}
\subsubsection{FP16 operator support}
Leveraging the half-precision vector multiply-add instruction (VFMULAH16) of the MT-3000, we implemented a high-performance FP16 operator library, which delivers approximately twice the peak performance of FP32 at the same frequency. A mixed-precision strategy of "FP16 storage with FP32 accumulation" is adopted: critical normalization and reduction operations (such as Softmax and RMSNorm) are computed at FP32 intermediate precision to prevent numerical underflow. The conversion between FP16 and FP32 is efficiently handled by native vector instructions with negligible overhead. For GEMM, we optimized the parameter configuration while maintaining the FP32 kernel structure to reduce intermediate data movement and enhance bandwidth utilization.
\subsection{Scheduling Strategy for Computational Graphs}
\subsubsection{Computational Graph Partitioning and Fusion Rules}
Based on the operator graph in Fig.~\ref{fig_4}, we propose a density-driven fusion method: fusion units are constructed in a ``[low-density]--high-density--[low-density]'' pattern, with each unit containing at least one high-density operator, which can incorporate multiple low-density operators either before or after it. The fusion boundaries are determined by two constraints: first, the data dependency distance—ensuring that data production and consumption form a closed loop within the unit to eliminate off-chip memory access; second, On-chip memory capacity—statically setting an upper limit based on the remaining capacity of AM, SM, and GSM to prevent working set overflow. Regions 1, 3, 4, and 5 in Fig.~\ref{fig_4} serve as typical examples. Taking the fusion of Linear and RoPE as an example, we embed RoPE into the Linear kernel, allowing results to be consumed on-the-fly in registers or on-chip caches, thereby further reducing I/O operations and kernel launch overhead. Implementation details are provided in Algorithm~\ref{alg:linear_fusion}.
\begin{algorithm}[t]
\caption{Linear–RoPE Fusion (compact)}
\label{alg:linear_fusion}
\begin{algorithmic}[1]
\footnotesize
\State $\theta_{id}\gets\bigl((n_1+n_2)\bmod \mathrm{dim}\bigr)\ /\ 32\;\;\text{(floor)}$
\For{$\mathrm{col}=0$ \textbf{to} $N_1/32-1$}
  \State $\mathrm{VPU\_load}\;\theta[\theta_{id}{+}{+}] \rightarrow VR_{\theta}[\mathrm{col}]$
  \For{$\mathrm{row}=0$ \textbf{to} $M_1-1$}
    \State $\mathrm{SPU\_load}\bigl(\mathrm{row}{+}m_0{+}m_2{+}\mathrm{Pid}\cdot M_2{+}m_1\bigr)\rightarrow R_{\mathrm{row}}$
    \State $\mathrm{Broadcast}\;R_{\mathrm{row}}\rightarrow VR_{\mathrm{row}}$
    \For{$\mathrm{col}=0$ \textbf{to} $N_1/32-1$}
      \State $\mathrm{VEC\_NEG}\;VR_y[\mathrm{row}][\mathrm{col}]\rightarrow VR_{\mathrm{neg}}[\mathrm{col}]$
      \State $\mathrm{bale2lh}\bigl(VR_y[\mathrm{row}][\mathrm{col}],\,VR_{\mathrm{neg}}[\mathrm{col}]\bigr)\rightarrow VR_{\mathrm{mix}}[\mathrm{col}]$
      \State $\mathrm{vec\_muli}\;VR_{\theta}[\mathrm{col}],\,VR_{\mathrm{row}}\rightarrow VR_{\mathrm{IP}}[\mathrm{col}]$
      \State $\mathrm{vm\_sinf32\_u35}\;VR_{\mathrm{IP}}[\mathrm{col}]\rightarrow VR_{\mathrm{sin}}[\mathrm{col}]$
      \State $\mathrm{vm\_cosf32\_u35}\;VR_{\mathrm{IP}}[\mathrm{col}]\rightarrow VR_{\mathrm{cos}}[\mathrm{col}]$
    \EndFor
    \For{$\mathrm{col}=0$ \textbf{to} $N_1/32 - 1$} \Comment{loop unrolled}
  \State $\mathrm{muli}\;VR_y[\mathrm{row}][\mathrm{col}],\,VR_{\mathrm{cos}}[\mathrm{col}]
          \rightarrow VR_{\mathrm{res}}[\mathrm{col}]$
\EndFor
\For{$\mathrm{col}=0$ \textbf{to} $N_1/32 - 1$} \Comment{loop unrolled}
  \State $\mathrm{Mula}\;VR_{\mathrm{mix}}[\mathrm{col}],\,VR_{\mathrm{sin}}[\mathrm{col}],\,VR_{\mathrm{res}}[\mathrm{col}]
          \rightarrow VR_{\mathrm{res}}[\mathrm{col}]$
\EndFor
  \EndFor
  \State $\mathrm{VPU\_store}\;VR_{\mathrm{res}}[0\!:\!N_2/32]\rightarrow Y_2[m_1{+}\mathrm{row}][n_1\!:\!n_1{+}N_1]$
\EndFor
\end{algorithmic}
\end{algorithm}

The algorithm uses VPU\_load to load the RoPE rotation angle \( \theta \) into the vector register \( \text{VR}\_{\theta} \); the scalar register \( R\_row \) stores the row index, which is broadcast to the vector \( \text{VR}\_{row} \). It combines the instructions vm\_sinf32\_u35 and vm\_cosf32\_u35 to implement complex number rotation, and \( \text{VR}\_{mix} \) is generated through VEC\_NEG and bale2lh instructions.
\subsubsection{MT Attention: Staged Pipeline-Based Fusion Optimization}
\begin{figure}[!tb]
    \centering
    \includegraphics[width=0.8\columnwidth]{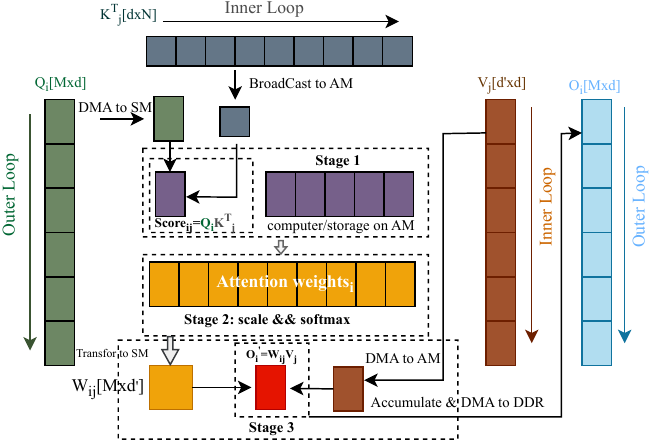}
    \caption{MT Attention: Staged Pipeline-Based Fusion Optimization Strategy}
    \label{fig_7}
\end{figure}

Focusing on the matrix dimensions of $Q_i$ and $K_i$ in attention mechanisms, we propose a hardware-aware phased fusion optimization strategy. For small sequence lengths ($S$), intermediate attention scores are cached in on-chip memory through \textit{operator fusion} to eliminate redundant memory accesses. When $S$ exceeds the SM capacity threshold, we dynamically adjust row-block sizes ($M$ rows) using an outer product method and utilize \textit{GSM} to cache attention scores, enabling block reuse and deep computational synergy. The optimization workflow, shown in Fig.~\ref{fig_7}, employs row-granular scheduling per attention head, with computation for a single head formulated as shown in Eq.~\ref{eq6}.

\begin{equation}
\label{eq6}
\text{Attention}(Q_i, K_i, V_i) = \text{softmax}\left(\frac{Q_i K_i^T}{\sqrt{d}}\right) V_i
\end{equation}
\par
It is divided into three stages:

\textbf{Stage 1}: Partition \( Q_i \) into row blocks (\( M \) rows) and load them into SM. Load \( K_i^T \) into AM in column blocks (\( N \) columns). After iteration, obtain the Attention Scores, with some stored in GSM and the rest in AM.

\begin{figure*}[htbp]
    \centering
    \includegraphics[width=0.7\textwidth]{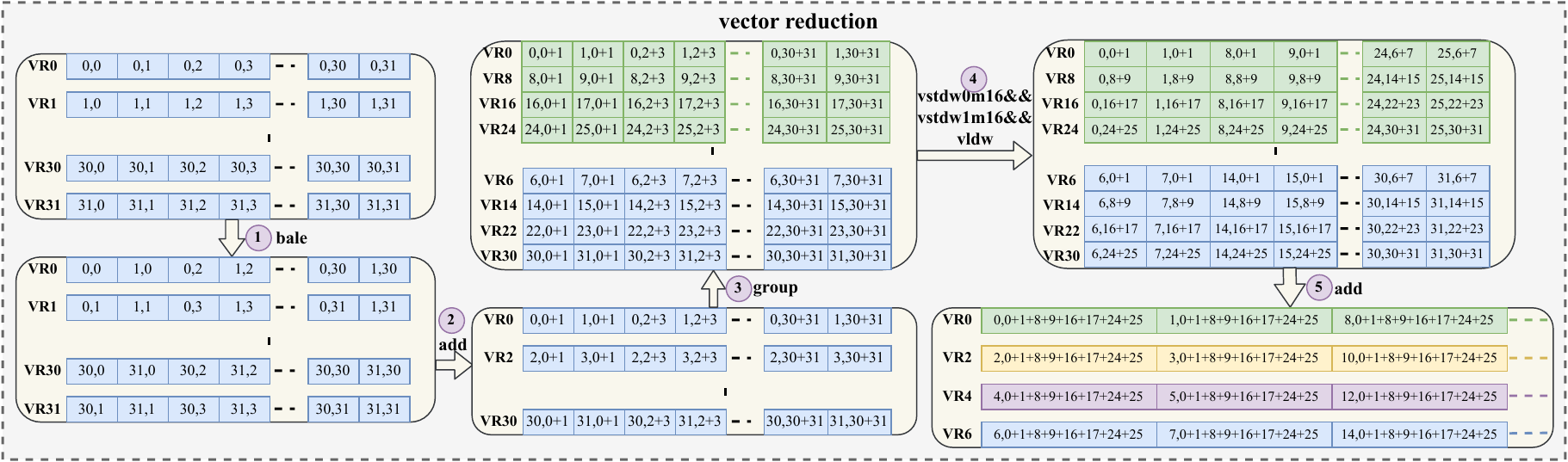}
    \caption{Hardware-Aware Reduction Algorithm: (1) Use the \texttt{bale} instruction to uniformly pack the high and low 32 bits of the register; (2) Add every two registers as a group; (3) Group the 16 added registers; (4) Use shuffle operations \texttt{vstdw0m16} and \texttt{vstdw1m16} to move the data from each group of registers to AM, and then load it back into the registers; (5) Add each group of registers to obtain the partial sum of each vector.}
    \label{fig_8}
\end{figure*}

\textbf{Stage 2}: Scale the \( M \) rows of Attention Scores obtained in Stage 1 and perform the softmax operation to get Attention Weights. Accelerate the Softmax using a hardware-customized vector reduction algorithm, with detailed steps shown in Fig.~\ref{fig_8}. Repeat steps 4-5 to ultimately obtain the reduction result of the sum inside 32 floating-point vectors.

\textbf{Stage 3}: Perform GEMM operations on the \( M \) rows of Attention Weights and \( V_i \) to obtain \( M \) rows of the final output \( O_i \) for a single head.

By iterating through these three stages multiple times, we can obtain the final result for a single head. Through this scheduling approach, the transmission of Attention Scores and Attention Weights is completely hidden, reducing I/O overhead and lowering the on-chip space complexity to \( O(M) \).
\par
The proposed MT Attention method and Flash Attention exhibit significant differences in core I/O behavior on the MT-3000 platform. Flash Attention requires keeping the K/V blocks fixed in AM and repeatedly loads Q blocks and output O blocks via DMA. In contrast, MT Attention keeps the Q/O blocks fixed in on-chip memory, executes in parallel across multiple DSPs, and iteratively loads K/V blocks via broadcasting. Assuming each loaded K/V block contains \(M_c\) rows and each Q/O block contains \(M_r\) rows.
\begin{table}[htbp]
\centering
\caption{I/O operation count comparison between MT Attention and Flash Attention}
\label{tab:io_comparison}
\begin{tabular}{@{}ccc@{}}
\toprule
I/O Type & Flash Attention & MT Attention \\
\midrule
Q DMA        & \(S/M_c\)            & \(1\) \\
O DMA        & \(2S/M_c\)           & \(1\) \\
K Broadcast  & \(1\)                & \(S/(M_r\times 24)\) \\
V Broadcast  & \(1\)                & \(S/(M_r\times 24)\) \\
\bottomrule
\end{tabular}
\end{table}
The I/O complexity comparison between the two methods is shown in Table~\ref{tab:io_comparison}. By slightly increasing the number of broadcast operations, MT Attention significantly reduces the number of DMA operations. Since the actual broadcast bandwidth is higher than that of DMA (with broadcast latency being about 0.9 times that of DMA), this strategy effectively reduces the overall I/O overhead and latency in Attention computation.
\subsubsection{Unified Kernel Scheduling and Load Balancing}
This study employs the Hthreads heterogeneous threading library to implement parallel scheduling for the intra-cluster DSP acceleration array. In the current scheduling mechanism, the total execution time for a single kernel is defined as:
\begin{equation}
T_{\text{single}} = T_{\text{create\_group}} + T_{\text{launch\_group}} + T_{\text{exec\_group}},
\end{equation}
where $T_{\text{create\_group}}$ is the thread group creation time, $T_{\text{launch\_group}}$ is the thread group launch time, and $T_{\text{exec\_group}}$ is the actual execution time. For a computation graph containing $N$ operators where each requires launching a separate thread group, the total overhead becomes:
\begin{equation}
T_{\text{total}} = N \cdot T_{\text{single}}.
\end{equation}
\par
Under the MHA mechanism, a short input sequence length readily induces load imbalance across multiple DSPs. This problem originates from two primary factors: the coarse task granularity in per-head scheduling mode impedes effective workload partitioning and leads to low resource utilization, while the need for independent kernel initialization and termination per attention head introduces substantial repeated scheduling overhead (\(T_{\text{create\_group}} + T_{\text{launch\_group}}\)). To address these issues, we propose a unified kernel scheduling strategy that operates at the granularity of the complete computation graph. This approach performs thread group creation and launch in a single operation, effectively eliminating multiple kernel launch overheads while enabling fine-grained task distribution across attention heads, thereby significantly alleviating load imbalance under short-sequence input scenarios.
\subsection{Adaptive Parallel Scheduling}
\subsubsection{Selective Batching}
To enhance system throughput under bandwidth constraints, THInfer adopts a \textit{selective batching} strategy. This approach is based on weight sharing analysis: except for the \textbf{Q}, \textbf{K}, \textbf{V} matrices in the Attention mechanism, other weights (such as those in Linear and LayerNorm layers) can be fully shared within a batch. Thus, traditional batching is applied to these compute-intensive operators. In contrast, the Attention component is processed on a per-task basis due to its I/O-intensive nature and dynamically independent KV cache. This hybrid strategy aims to meticulously balance the trade-off between latency and throughput. 
\par
Let the effective batch size be \(B\), the input sequence length be \(S\), and the output sequence length be \(N\). 
\begin{itemize}
    \item \textit{Latency} (\(T_{\text{latency}}\)) is defined as the total time required to process the prompt and generate all \(B \times N\) tokens.
    \item \textit{Throughput} is defined as \(B \times N / T_{\text{latency}}\) (tokens/s).
\end{itemize}
\subsubsection{Resource-Aware P-B-D Three-Level Synchronous Pipeline}
\begin{figure}[htbp]
\centering
\includegraphics[width=\columnwidth]{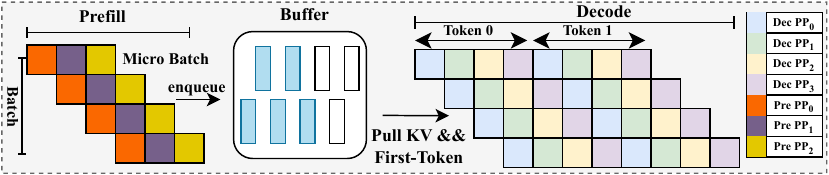}
\caption{P-B-D Three-Level Synchronous Pipeline}
\label{fig:pipeline}
\end{figure}

During inference, the Prefill and Decode stages exhibit significant bottleneck heterogeneity. While a fully asynchronous decoupling strategy could improve system throughput, it may lead to KV cache accumulation and subsequent GPU out-of-memory (OOM) errors when Decode latency is high. To address this issue, THInfer designs a Prefill-Buffer-Decode (P-B-D) three-level synchronous pipeline (Fig.~\ref{fig:pipeline}), which achieves inter-stage decoupling and parallel execution at a micro-batch granularity. By incorporating a bounded buffer and a backpressure control mechanism, it restricts how far the Prefill stage can advance ahead, thus structurally avoiding unbounded cache growth issues similar to those in DistServe.

This pipeline divides the cluster into three logical pools based on functionality:
\begin{itemize}
    \item \textbf{Prefill Pool}: Handles incoming requests, performs prefill computation, and generates the first token along with its corresponding initial KV cache. This pool employs Data Parallelism (DP) and Pipeline Parallelism (PP) strategies to fully exploit GEMM computational throughput.
    
    \item \textbf{Buffer Pool}: Acts as an intermediate staging area, maintaining a bounded buffer with a capacity of \(\mathit{Buf}_{\max}\) on idle nodes, specifically for temporarily storing KV cache.
    
    \item \textbf{Decode Pool}: Synchronously fetches data at the micro-batch granularity, maps the corresponding KV cache in a pull-based manner, and employs Tensor Parallelism (TP), PP and DP strategies to distribute the storage and bandwidth pressure of model weights and KV cache.
\end{itemize}
\par
To ensure system stability, a bounded buffer and a backpressure control mechanism are introduced: when buffer utilization reaches its upper limit, backpressure will be applied to the Prefill stage, pausing the acceptance of new requests, thereby preventing KV cache overflow. The Decode stage pulls KV data on-demand upon readiness, effectively shortening the cache lifecycle and suppressing peak memory usage.

\subsubsection{Hybrid Parallel Strategy Based on Resource Constraints}
To address the challenge that the memory requirements of models with tens of billions of parameters far exceed the capacity of a single cluster, THInfer performs systematic modeling and configuration optimization of hybrid parallelism strategies based on hardware resource constraints. Let the total model parameter size be \( \mathit{Model\_size} \), the number of layers be \( L \), the maximum context length be \( S_{\max} \), the word embedding dimension be \( D_{\text{emb}} \), and the byte size of the data be \( D_{\text{size}} \). Let the Tensor Parallelism degree be \( TP \), the Pipeline Parallelism degree be \( PP \), the micro-batch size be \( B_{\text{micro}} \), and the Data Parallelism degree be \( DP \). The total batch size is \( B \).
\begin{itemize}
    \item \textbf{Memory Overhead Modeling}
    \par
    The per-cluster memory footprint during the Prefill stage is:
    \begin{equation}
    \small
\text{Params}_{\text{per\_cluster}} = \frac{\mathit{Model\_size}}{PP_P},
    \label{eq:params_prefill}
    \end{equation}    
    \begin{equation}
    \small
\text{KV}_{\text{per\_cluster}} = B_{\text{micro\_P}} \cdot S_{\max} \cdot 2 \cdot D_{\text{size}} \cdot D_{\text{emb}}.
    \label{eq:kv_prefill}
    \end{equation}
    \par
    The per-cluster memory footprint during the Decode stage is:
    \begin{equation}
    \small
\text{Params}_{\text{per\_cluster}} = \frac{\mathit{Model\_size}}{TP \cdot PP_D},
    \label{eq:params_decode}
    \end{equation}
    \begin{equation}
    \small
\text{KV}_{\text{per\_cluster}} = \frac{B_{\text{micro\_D}} \cdot S_{\max} \cdot 2 \cdot D_{\text{size}} \cdot D_{\text{emb}} \cdot L}{TP}.
    \label{eq:kv_decode}
    \end{equation}
    \par
    The constraint condition is:
    \begin{equation}
\text{Params}_{\text{per\_cluster}} + \text{KV}_{\text{per\_cluster}} + \text{Overhead} < 20\,\text{GB}.
    \end{equation}
    \item \textbf{Communication Overhead Modeling}
    \par
    The primary communication overhead comes from the All-Reduce operation within the TP group during the Decode stage:
    \begin{equation}
    \text{Comm}_{TP} = B \cdot 2 \cdot D_{\text{emb}} \cdot D_{\text{size}} \cdot (TP - 1),
    \end{equation}
    
    KV cache is asynchronously migrated using MPI non-blocking communication: upon completion of a single layer computation in the Prefill stage, the corresponding KV cache is transmitted asynchronously. The Prefill Pool can simultaneously proceed with the computation of the next layer, thereby overlapping communication with computation.

    \item \textbf{Performance Modeling and Parallel Configuration Search}
    \par
    The computation time for a single layer on a single cluster without TP is:
    \begin{equation}
    T_{\text{single\_clu}} = T_{\text{norm}} + T_{\text{self\_attention}} + T_{\text{norm}} + T_{\text{ffn}}.
    \end{equation}
    \par
    After applying TP, the single layer computation time becomes:
    \begin{equation}
    \small
    \begin{split}
    T_{\text{clu}} = & \; T_{\text{norm}} + \frac{T_{\text{self\_attention}}}{TP} + T_{\text{norm}} \\
    & + \frac{T_{\text{ffn}}}{TP} + 2 \cdot T_{\text{all\_reduce}} + 2 \cdot T_{\text{launch\_group}}.
    \end{split}
    \label{eq:clu_time}
    \end{equation}
    \par
    The benefit of TP depends on the trade-off between the reduction in computation time (primarily \( T_{\text{self\_attention}} \) and \( T_{\text{ffn}} \)) and the introduced additional overhead.
    To facilitate practical deployment, we systematically perform a parallel configuration search on the Decode size based on \( DP_D = 1 \): first, under the constraints of memory usage and communication overhead, we search for a \( TP_D \) that balances latency and throughput; second, we determine \( PP_D \) within the memory constraints, thereby obtaining the target batch size \( B = PP_D \times \text{micro\_batch} \); the size of the Buffer Pool \( B_p \) is determined by the KV cache size corresponding to batch size \( B \); the \( PP_P \) on the Prefill side is directly computed based on the minimum memory overhead constraint; under the condition that the Decode Pool occupies \( TP_D \times PP_D \) computing clusters, the cluster overhead of a parallel pool is:
    \begin{equation}
    \text{Clu}_{\text{pool}} = DP_P \cdot PP_P + TP_D \cdot PP_D + B_P.
        \end{equation}
    \par
    The overall throughput is approximately:
    \begin{equation}
    \text{Throughput} \propto \frac{\text{Clu}_{\text{total}}}{\text{Clu}_{\text{pool}}} \cdot \frac{1}{\max(T_p / DP_P, T_d)}.
    \end{equation}
    
    Thus, maximizing throughput is equivalent to minimizing:
    \begin{equation}
    \min_{DP_P} \left\{ \text{Clu}_{\text{pool}} \cdot \left( \max(T_p / DP_P, T_d) \right) \right\},
    \end{equation}
    where \( T_p \) and \( T_d \) represent the latency of the Prefill stage and Decode stage, respectively, when \( DP = 1 \). These can be estimated for a given batch size through simulation or stress testing. The optimal \( DP_P \) configuration that satisfies the memory and bandwidth constraints is then solved.
\end{itemize}

\section{Experiments}
\subsection{Experimental Setup}
\begin{itemize}
    \item \textbf{Hardware}: Experiments were conducted on MT-3000 nodes, with comparisons made against Tesla V100S-PCIE-32GB and NVIDIA A800 80GB PCIe.
    \item \textbf{Model}: Evaluation was performed using the Llama2 series of models, with parameter sizes ranging from 7B to 70B. Although other models were not tested, the techniques employed by the THInfer framework are equally applicable to other LLMs based on the Transformer architecture (such as Qwen, DeepSeek, etc.).
    \item \textbf{Workload}: The experiments focused on high-throughput text generation tasks. A synthetic prompt dataset, padded to a uniform length, was used, with each prompt requiring the generation of 128 tokens. Two prompt lengths were selected for testing: 512 and 1024. The evaluation metric was generation throughput.
    \item \textbf{Baseline}: The baseline systems included DeepSpeed-Inference's~\cite{aminabadi2022deepspeed} tensor parallelism scheme and Hugging Face Accelerate's~\cite{huggingface_tgi} pipeline parallelism scheme.
    \item \textbf{Implementation}: THInfer is developed in pure C++ with inline assembly, its codebase comprising roughly ten thousand lines.
\end{itemize}

\subsection{End-to-End Throughput Tests}
To comprehensively evaluate system performance, end-to-end throughput tests were conducted. The baseline performance of the hardware platforms involved in the comparison is shown in Table~\ref{tab:hardware_specs}.
\begin{table}[h]
\centering
\caption{Hardware Platform Performance Specifications}
\label{tab:hardware_specs}
\begin{tabular}{lcc}
\toprule
\textbf{Device} & \textbf{FP16 Performance} & \textbf{Bandwidth} \\
\midrule
MT-3000 & 32.4 TFLOPS & 120 GB/s \\
V100S-PCIE-32GB & 130 TFLOPS & 1134 GB/s \\
A800 80GB PCIe & 312 TFLOPS & 1935 GB/s \\
\bottomrule
\end{tabular}
\end{table}

To ensure a fair comparison, the experiments followed a peak-performance alignment principle: 8 MT-3000 devices were compared against 2 V100S cards, and 10 MT-3000 devices were compared against 1 A800 card. Since DeepSpeed and Accelerate demonstrated similar performance in a single-card environment, only DeepSpeed was used as the baseline for A800. To further investigate performance under bandwidth-aligned conditions, additional configurations of 18 and 16 MT-3000 devices were tested to match the bandwidth levels of 2×V100S and 1×A800, respectively. The experimental results are shown in Tables~\ref{tab:throughput_v100s} and ~\ref{tab:throughput_a800}.
\begin{table}[t]
\centering
\caption{Throughput Comparison: MT-3000 vs. V100S (Tokens/s)}
\label{tab:throughput_v100s}
\setlength{\tabcolsep}{4pt}
\renewcommand{\arraystretch}{1.05}
\small
\resizebox{\linewidth}{!}{%
\begin{tabular}{ccccccccc}
\toprule
 & \multicolumn{2}{c}{\textbf{Accelerate}} & \multicolumn{2}{c}{\textbf{DeepSpeed}} & \multicolumn{2}{c}{\textbf{THInfer (Peak-Aligned)}} & \multicolumn{2}{c}{\textbf{THInfer (BW-Aligned)}} \\
\cmidrule(lr){2-3}\cmidrule(lr){4-5}\cmidrule(lr){6-7}\cmidrule(lr){8-9}
 & 512 & 1024 & 512 & 1024 & 512 & 1024 & 512 & 1024 \\
\midrule
7B  & 323 & 173 & 481 & 263 & \textbf{781} & \textbf{456} & {1755} & 1026 \\
13B & 168 &  93 & \textbf{273} & 145 & 241 & \textbf{169} & {543} & 381 \\
30B & 1.9 & 0.82 & 7.69 & 4.07 &  \textbf{81} &  \textbf{46} & {186} & 105 \\
70B & \textsc{oom} & \textsc{oom} & \textsc{oom} & \textsc{oom} & \textbf{64} & \textbf{41} & {129} & 81 \\
\bottomrule
\end{tabular}%
}
\end{table}
\begin{table}[t]
\centering
\caption{Throughput Comparison: MT-3000 vs. A800 (Tokens/s)}
\label{tab:throughput_a800}
\setlength{\tabcolsep}{6pt}
\renewcommand{\arraystretch}{1.05}
\small
\resizebox{\linewidth}{!}{%
\begin{tabular}{ccccccc}
\toprule
 & \multicolumn{2}{c}{\textbf{DeepSpeed}} & \multicolumn{2}{c}{\textbf{THInfer (Peak-Aligned)}} & \multicolumn{2}{c}{\textbf{THInfer (BW-Aligned)}} \\
\cmidrule(lr){2-3}\cmidrule(lr){4-5}\cmidrule(lr){6-7}
 & 512 & 1024 & 512 & 1024 & 512 & 1024 \\
\midrule
7B  & 584 & 310 &  \textbf{975} & \textbf{570} &  {1560} & 912 \\
13B & \textbf{345} & 184 &  301 & \textbf{211} &   {483} & 339 \\
30B & \textbf{116} &  60 &  105 &  \textbf{61} &   {162} &  89 \\
70B & \textsc{oom} & \textsc{oom} &  \textbf{64} &  \textbf{41} &  {129} &  81 \\
\bottomrule
\end{tabular}%
}
\end{table}
\par
The experimental results demonstrate that under the peak-performance-aligned configuration, MT-3000 achieves significant advantages or delivers comparable performance across various model sizes and input lengths. For the \textbf{7B model}, MT-3000 achieves a 62\%--73\% improvement over V100S (using DeepSpeed) and a 67\%--84\% improvement over A800, underscoring its capability to efficiently leverage GEMM computation during the prefill phase via handcrafted kernels and operator fusion. In the \textbf{13B model}, MT-3000 outperforms competing systems on long-sequence (1024) tasks, indicating that its computational advantage is more pronounced in compute-intensive scenarios. For the \textbf{30B and 70B models}, the performance lead of MT-3000 further expands. Notably, in the 70B model, GPU-based solutions encountered out-of-memory (OOM) errors, whereas MT-3000 sustained efficient inference through hybrid parallelism and memory expansion mechanisms, demonstrating exceptional scalability.
\par
Under the bandwidth-aligned configuration, the system throughput shows near-linear improvement (e.g., approximately 2.25× for 7B, 13B, and 30B models, and about 2× for the 70B model), indicating that aggregated DDR bandwidth effectively alleviates the KV cache read/write bottleneck during the decoding phase. Combined with deep computational graph fusion, selective batching, and the P-B-D three-level pipeline mechanism, THInfer successfully synergizes the high computational efficiency of the prefill phase with the bandwidth scalability of the decoding phase, achieving a seamless transition from core-level compute optimization to system-level I/O expansion.
\par
In summary, THInfer excels in long-context and large-model scenarios, with performance advantages becoming more significant as input length and model size increase. This validates that its hardware-software co-design effectively addresses various bottlenecks in Transformer inference. The bandwidth-aligned experiments further confirm the system's strong scalability, providing a stable and efficient solution for inference of billion-parameter models.
\subsection{Ablation Study}
To evaluate the effectiveness of each optimization strategy, we conducted an ablation study. The experimental setup was as follows: a prompt length of 512, generating 128 tokens, using a 12-cluster configuration. The baseline version (A0) did not employ any optimizations and used only single-node deployment with multi-node data parallelism; we then progressively introduced operator optimization (A1), computational graph-level algorithm optimization (A2), batching (A3 series), and the P–B–D three-stage pipeline scheduling (A4). A3 and A4 correspond to system-level optimizations, and the experimental results are shown in Table~\ref{tab:ablation}.
\begin{table}[t]
\centering
\caption{Throughput comparison under different optimization strategies (tokens/s).}
\label{tab:ablation}
\setlength{\tabcolsep}{4pt}
\renewcommand{\arraystretch}{1.08}
\small
\resizebox{\linewidth}{!}{%
\begin{tabular}{lccccp{0.45\linewidth}ccc}
\toprule
\textbf{Group} & \textbf{Operator Opt.} & \textbf{Graph Sched.} & \textbf{Batching} & \textbf{P–B-D} & \textbf{Key Features} & \textbf{7B} & \textbf{13B} & \textbf{30B} \\
\midrule
A0   & $\times$ & $\times$ & $\times$ & $\times$ & Single-node deployment with multi-node data parallelism & 1.02 & 0.21 & 0.07 \\
A1   & \checkmark & $\times$ & $\times$ & $\times$ & Enable FP16 high-performance operators & 21 & 5.80 & 2.47 \\
A2   & \checkmark & \checkmark & $\times$ & $\times$ & Enable computation graph optimization & 26 & 7.22 & 3.02 \\
A3-1 & \checkmark & \checkmark & \checkmark & $\times$ & Enable selective batching (PP only) & 124 & 40 & 14 \\
A3-2 & \checkmark & \checkmark & \checkmark & $\times$ & Enable selective batching (TP + PP) & 312 & 63 & 27 \\
A4   & \checkmark & \checkmark & \checkmark & \checkmark & Enable P–B–D three-stage pipeline & \textbf{321} & \textbf{101} & \textbf{34} \\
\bottomrule
\end{tabular}%
}
\end{table}
\par
As shown in Table~\ref{tab:ablation}, the ablation study comprehensively validates the performance contribution of each optimization module within THInfer. From A0 to A4, all models exhibit a stepwise increase in throughput, with models of different scales showing distinct responses to the optimization strategies. Operator optimization (A1), leveraging FP16 handcrafted assembly kernels to fully utilize hardware compute capabilities, increased the throughput of the 30B model from 0.07 token/s to 2.47 token/s, a 35-fold improvement; the benefit of this optimization amplifies with increasing model scale, indicating that large-scale GEMM computations are more reliant on low-level compute optimizations. Computational graph scheduling optimization (A2) further provided a 22\% performance gain (30B model: 2.47 $\rightarrow$ 3.02 token/s), primarily achieved by fusing operators like MHA and RoPE to reduce redundant memory access and kernel launch overhead.
\par
Batching and parallelism strategies (A3 series) yielded significant improvements. Using only selective batching with PP (A3-1) achieved a throughput of 14 token/s for the 30B model, an approximately 4.6x improvement over A2. Introducing tensor parallelism (A3-2) further increased throughput for all models, but the magnitude of gain varied significantly: the 7B model increased from 124 to 312 token/s (+152\%), while the 13B and 30B models grew by approximately 58\% and 93
\% respectively, suggesting that as model scale increases, All-Reduce communication gradually becomes a scaling bottleneck.
\par
The P-B-D three-level pipeline scheduling (A4) delivered additional performance improvements across all model sizes, with the magnitude of enhancement becoming more substantial as model scale increased: a 3\% improvement for the 7B model (312 $\rightarrow$ 321 tokens/s), a 60\% increase for the 13B model (63 $\rightarrow$ 101 tokens/s), and a 26\% gain for the 30B model (27 $\rightarrow$ 34 tokens/s). These results indicate that the P-B-D three-level pipeline effectively optimizes resource scheduling and data transfer efficiency, proving particularly advantageous for medium to large-scale models characterized by high decoding load and frequent KV cache transfers.
\subsection{Effectiveness of the Linear Operator}
This section tests the operational efficiency of the core LLM operator, Linear, on a single cluster of MT-3000. The experiment covers various scales, with sequence length (\(S\)) ranging from 1024 to 8192, input dimension (\textit{Input dim}) fixed at 4096, and output dimension (\textit{Output dim}) ranging from 128 to 4096. The test results are shown in Table~\ref{tab:linear_latency_fp32_fp16}.


\begin{table}[ht]
\centering
\caption{MT-3000 Single-Cluster Linear Operator Latency Testing Results (FP32/FP16, Unit: ms)}
    \label{tab:linear_latency_fp32_fp16}
\scriptsize
\setlength{\tabcolsep}{3.2pt}
\resizebox{\linewidth}{!}{%
\begin{tabular}{c|c|ccccccc}
\toprule
\textbf{Precision} & \textbf{Output dim} & \textbf{256} & \textbf{512} & \textbf{768} & \textbf{1024} & \textbf{2048} & \textbf{3072} & \textbf{4096} \\
\midrule
\multirow{4}{*}{FP32}
 & S=1k & 1.12 & 1.80 & 2.78 & 3.52 & 5.79 & 7.93 & 10.04 \\
 & S=2k & 1.74 & 3.06 & 4.23 & 5.06 & 8.20 & 11.51 & 14.66 \\
 & S=4k & 3.20 & 4.87 & 8.10 & 9.14 & 15.45 & 21.51 & 27.76 \\
 & S=8k & 6.23 & 9.50 & 13.95 & 17.23 & 29.28 & 41.45 & 53.77 \\
\midrule
\multirow{4}{*}{FP16}
 & S=1k & 0.52 & 0.64 & 0.90 & 1.62 & 2.03 & 2.93 & 3.88 \\
 & S=2k & 0.66 & 1.05 & 1.45 & 1.95 & 3.38 & 5.05 & 6.53 \\
 & S=4k & 1.59 & 2.19 & 3.28 & 3.62 & 6.80 & 9.54 & 12.52 \\
 & S=8k & 3.05 & 4.45 & 5.81 & 7.67 & 12.93 & 18.47 & 24.17 \\
\bottomrule
\end{tabular}%
}
\end{table}

The performance calculation is shown in Eq.~\eqref{eq7}.
\begin{equation}
\label{eq7}
Per = \frac{S}{1024} \times \frac{Input dim}{1024} \times \frac{Output dim}{1024} \times \frac{1000}{times} \text{ GFLOPS}.
\end{equation}
\begin{figure}[t]
  \centering
  \begin{minipage}[t]{0.49\linewidth}
    \centering
    \includegraphics[width=\linewidth]{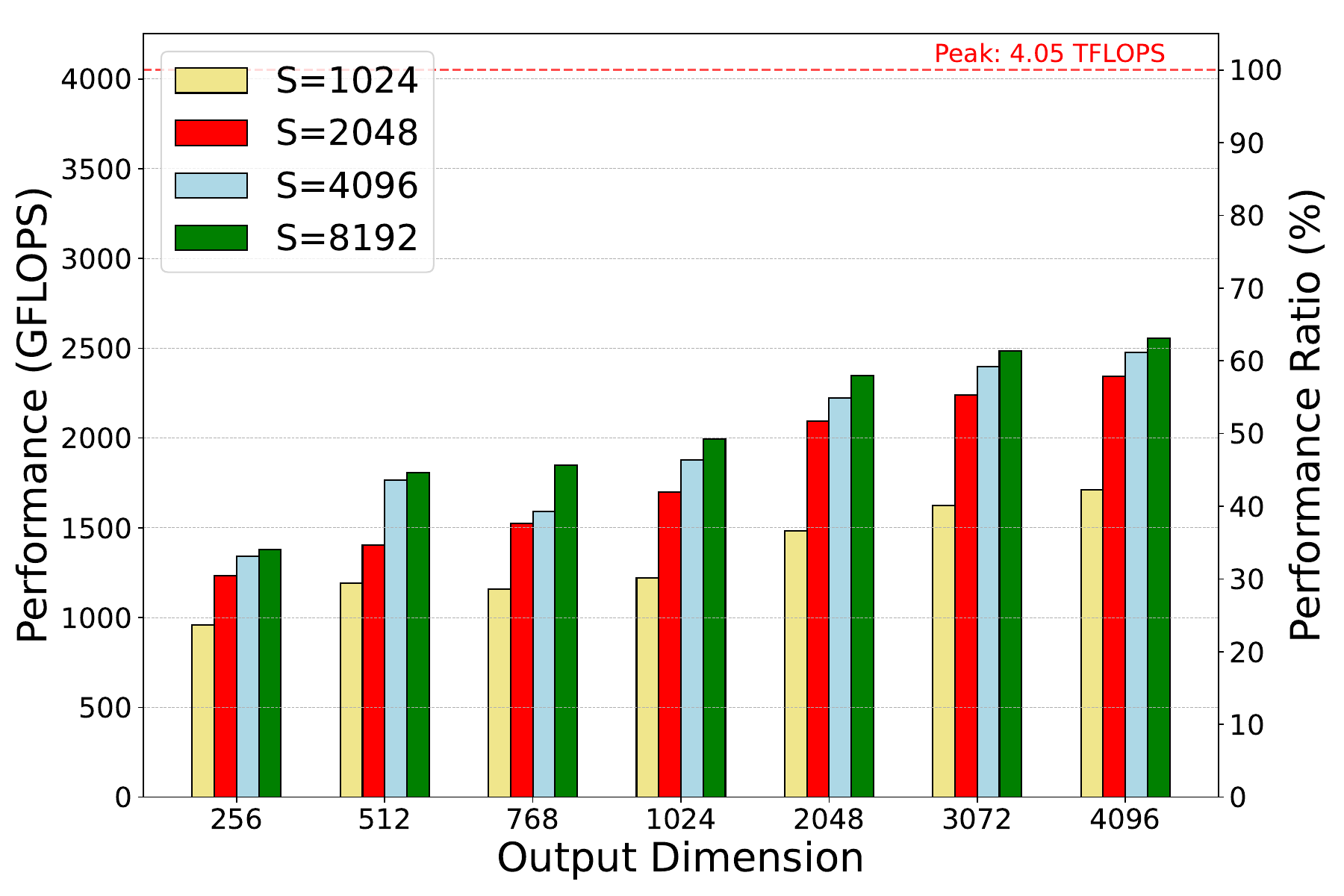}\\[2pt]
  \end{minipage}\hfill%
  \centering
  \begin{minipage}[t]{0.49\linewidth}
    \centering
    \includegraphics[width=\linewidth]{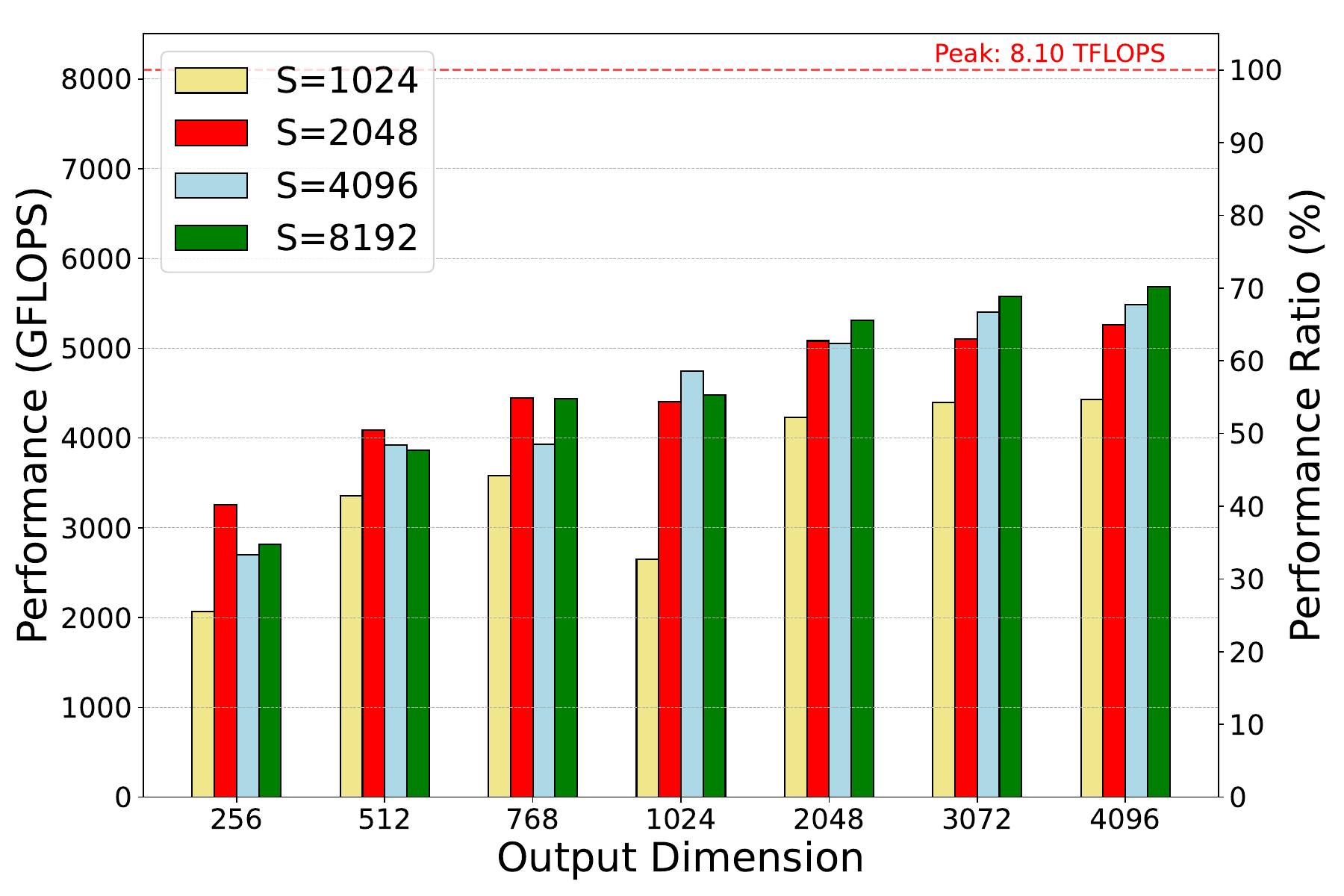}\\[2pt]
  \end{minipage}\hfill%
  \caption{Performance utilization and theoretical peak analysis for the Linear operator on MT-3000.}
  \label{fig11}
\end{figure}
\par
Based on the performance evaluation of the MT-3000 single-cluster Linear operator (Fig.~\ref{fig11}), the computational utilization increases with larger computational scales, primarily due to improved computational density and parallel efficiency from large-scale. This allows the VLIW-SIMD computing units to operate near saturation, effectively hiding memory access latency. The FP16 precision significantly enhances computational efficiency: under the configuration of \texttt{Input dim=4096}, \texttt{Output dim=4096}, and \texttt{S=8k}, the FP16 implementation achieves \textbf{5686.34 GFLOPS}, reaching \textbf{70.20\%} of the theoretical peak performance. This represents a \textbf{122\%} improvement compared to the FP32 performance of \textbf{2556.05 GFLOPS} (\textbf{63.11\%} utilization). The advantage stems from native hardware support for FP16 multiply-accumulate instructions (\texttt{VFMULAH16}) and reduced data volume, which optimizes memory access and alleviates bandwidth pressure on the data stream. The experiments also validate the effectiveness of the \texttt{DDR-GSM-AM} dataflow design and hand-tuned kernel design, which maximize vector register reuse through strategies such as loop unrolling and vector broadcasting. The peak utilization rate shows that the best-case scenario (FP16, \texttt{Input dim=4096}, \texttt{Output dim=4096}, \texttt{S=8k}) achieves \textbf{70.20\%} utilization, with the remaining bottleneck primarily attributed to DDR bandwidth limitations.
\subsection{Scaling Dot-Product Attention Operator Optimization Test}
This section evaluates the MT Attention optimization strategy for the scaled dot-product attention operator. Table~\ref{tab:attn_opt_latency} compares the performance before and after optimization.
\begin{table}[t]
\centering
\caption{Latency comparison before/after attention optimizations for scaled dot-product attention (batch size = 4, head\_dim = 128; unit: ms).}
\label{tab:attn_opt_latency}
\setlength{\tabcolsep}{4pt}
\renewcommand{\arraystretch}{1.08}
\small
\resizebox{\linewidth}{!}{%
\begin{tabular}{llcccccccc}
\toprule
\textbf{Method} & \textbf{Head} & \textbf{512} & \textbf{1024} & \textbf{1536} & \textbf{2048} & \textbf{2560} & \textbf{3072} & \textbf{3584} & \textbf{4096} \\
\midrule
Before Optimization              & Head=32 & 29.20 & 67.54 & 81.77 & 183.45 & 247.50 & 272.79 & 422.49 & 527.43 \\
                                 & Head=64 & 57.98 & 137.46 & 161.18 & 367.72 & 493.82 & 542.11 & 842.89 & 1058.14 \\
\midrule
After \textbf{MT Attention}       & Head=32 &  9.72 &  25.13 &  32.45 &  79.00 & 107.61 & 119.39 & 191.98 & 248.26 \\
                                 & Head=64 & 18.20 &  49.29 &  63.20 & 156.56 & 212.08 & 236.59 & 381.52 & 496.23 \\
\midrule
After \textbf{Flash Attention}    & Head=32 & 14.02 &  39.21 &  52.37 & 120.51 & 156.03 & 162.15 & 255.17 & 320.74 \\
                                 & Head=64 & 27.69 &  77.40 & 103.96 & 240.31 & 311.34 & 323.74 & 509.70 & 640.81 \\
\bottomrule
\end{tabular}%
}
\end{table}
\begin{figure}[h]
\centering
\includegraphics[width=\linewidth]{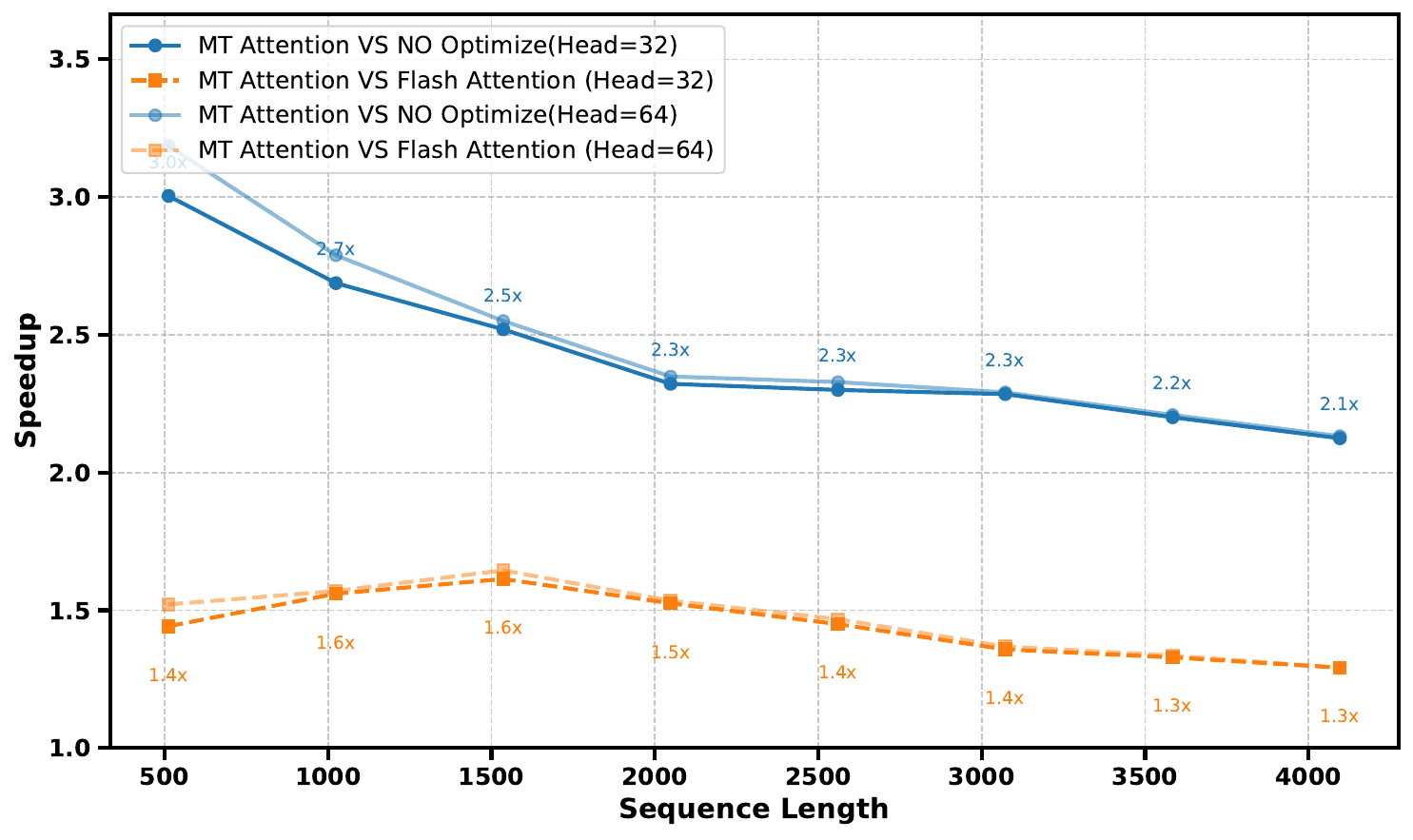}
\caption{Speedup of MT Attention on MT-3000}
\label{fig:attention_speedup}
\end{figure}
\par
The experimental results indicate that the phased fusion optimization of MT Attention significantly reduces the latency of the Scaled Dot-Product Attention operator. In long-sequence scenarios (e.g., S=4096, Head=64), performance improves by over 53\%, achieving a speedup of 2.5--3.5 times compared to the unoptimized version. Compared to FlashAttention, MT Attention is better adapted to the MT-3000 architecture, reducing latency by 26\%--36\% in most tests. Its advantage stems from effectively reducing DMA transfers and redundant I/O through broadcast mechanisms and a phased pipeline, fully leveraging the parallel potential of the heterogeneous many-core architecture. This optimization, synergizing with the system-level pipeline design, further validates the critical role of hardware-software co-design in enhancing the inference performance of large models.
\section{Conclusion}
This paper presents THInfer, an LLM inference framework designed for the Tianhe New-Generation Supercomputer. It effectively addresses the key challenges of deploying models with tens of billions of parameters on a many-core architecture with limited memory bandwidth and distributed memory hierarchy. Through hardware-software co-design and cross-layer optimization, three core contributions are achieved: a hand-optimized VLIW-SIMD operator library with FP16 support, boosting computational efficiency to 70.2\% of the theoretical peak; a density-driven operator fusion strategy and the MT Attention mechanism, significantly reducing I/O and kernel scheduling overhead; and a P–B–D pipeline with hybrid parallelism, enabling efficient decoupling of decoding and prefill phases while supporting scalable and stable inference even on 70B models. Experiments demonstrate that THInfer significantly outperforms GPU-based baseline systems in throughput, validating its effectiveness, scalability, and robustness under practical workloads. This work provides an essential technical pathway and practical foundation for efficient LLM inference on heterogeneous high-performance architectures.
\section*{Acknowledgment}
This work was supported by the National Key R\&D Program of China under Grant 2023YFB3001900, as well as the National Natural Science Foundation of China (NSFC) under Grants U23B2020 and 62322201.

\bibliographystyle{IEEEtran}

\bibliography{ref-all}

@article{vaswani2017attention,
  title={Attention is all you need},
  author={Vaswani, A},
  journal={Advances in Neural Information Processing Systems},
  year={2017}
}

@inproceedings{aminabadi2022deepspeed,
  title={Deepspeed-inference: enabling efficient inference of transformer models at unprecedented scale},
  author={Aminabadi, Reza Yazdani and Rajbhandari, Samyam and Awan, Ammar Ahmad and Li, Cheng and Li, Du and Zheng, Elton and Ruwase, Olatunji and Smith, Shaden and Zhang, Minjia and Rasley, Jeff and others},
  booktitle={SC22: International Conference for High Performance Computing, Networking, Storage and Analysis},
  pages={1--15},
  year={2022},
  organization={IEEE}
}

@article{ma2023llm,
  title={Llm-pruner: On the structural pruning of large language models},
  author={Ma, Xinyin and Fang, Gongfan and Wang, Xinchao},
  journal={Advances in neural information processing systems},
  volume={36},
  pages={21702--21720},
  year={2023}
}

@inproceedings{frantar2023sparsegpt,
  title={Sparsegpt: Massive language models can be accurately pruned in one-shot},
  author={Frantar, Elias and Alistarh, Dan},
  booktitle={International Conference on Machine Learning},
  pages={10323--10337},
  year={2023},
  organization={PMLR}
}

@article{wang2020linformer,
  title={Linformer: Self-attention with linear complexity},
  author={Wang, Sinong and Li, Belinda Z and Khabsa, Madian and Fang, Han and Ma, Hao},
  journal={arXiv preprint arXiv:2006.04768},
  year={2020}
}

@article{lin2024awq,
  title={AWQ: Activation-aware Weight Quantization for On-Device LLM Compression and Acceleration},
  author={Lin, Ji and Tang, Jiaming and Tang, Haotian and Yang, Shang and Chen, Wei-Ming and Wang, Wei-Chen and Xiao, Guangxuan and Dang, Xingyu and Gan, Chuang and Han, Song},
  journal={Proceedings of Machine Learning and Systems},
  volume={6},
  pages={87--100},
  year={2024}
}

@article{dao2022flashattention,
  title={Flashattention: Fast and memory-efficient exact attention with io-awareness},
  author={Dao, Tri and Fu, Dan and Ermon, Stefano and Rudra, Atri and R{\'e}, Christopher},
  journal={Advances in Neural Information Processing Systems},
  volume={35},
  pages={16344--16359},
  year={2022}
}

@inproceedings{yu2024optimizing,
  title={Optimizing General Matrix Multiplications on Modern Multi-core DSPs},
  author={Yu, Kainan and Qi, Xinxin and Zhang, Peng and Fang, Jianbin and Dong, Dezun and Wang, Ruibo and Tang, Tao and Huang, Chun and Che, Yonggang and Wang, Zheng},
  booktitle={2024 IEEE International Parallel and Distributed Processing Symposium (IPDPS)},
  pages={964--975},
  year={2024},
  organization={IEEE}
}

@article{lu2022mt,
  title={MT-3000: a heterogeneous multi-zone processor for HPC},
  author={Lu, Kai and Wang, Yaohua and Guo, Yang and Huang, Chun and Liu, Sheng and Wang, Ruibo and Fang, Jianbin and Tang, Tao and Chen, Zhaoyun and Liu, Biwei and others},
  journal={CCF Transactions on High Performance Computing},
  volume={4},
  number={2},
  pages={150--164},
  year={2022},
  publisher={Springer}
}

@article{touvron2023llama,
  title={Llama: Open and efficient foundation language models},
  author={Touvron, Hugo and Lavril, Thibaut and Izacard, Gautier and Martinet, Xavier and Lachaux, Marie-Anne and Lacroix, Timoth{\'e}e and Rozi{\`e}re, Baptiste and Goyal, Naman and Hambro, Eric and Azhar, Faisal and others},
  journal={arXiv preprint arXiv:2302.13971},
  year={2023}
}

@article{touvron2023llama2,
  title={Llama 2: Open foundation and fine-tuned chat models},
  author={Touvron, Hugo and Martin, Louis and Stone, Kevin and Albert, Peter and Almahairi, Amjad and Babaei, Yasmine and Bashlykov, Nikolay and Batra, Soumya and Bhargava, Prajjwal and Bhosale, Shruti and others},
  journal={arXiv preprint arXiv:2307.09288},
  year={2023}
}

@article{dubey2024llama,
  title={The llama 3 herd of models},
  author={Dubey, Abhimanyu and Jauhri, Abhinav and Pandey, Abhinav and Kadian, Abhishek and Al-Dahle, Ahmad and Letman, Aiesha and Mathur, Akhil and Schelten, Alan and Yang, Amy and Fan, Angela and others},
  journal={arXiv preprint arXiv:2407.21783},
  year={2024}
}

@article{bai2023qwen,
  title={Qwen technical report},
  author={Bai, Jinze and Bai, Shuai and Chu, Yunfei and Cui, Zeyu and Dang, Kai and Deng, Xiaodong and Fan, Yang and Ge, Wenbin and Han, Yu and Huang, Fei and others},
  journal={arXiv preprint arXiv:2309.16609},
  year={2023}
}

@article{sze2017efficient,
  title={Efficient processing of deep neural networks: A tutorial and survey},
  author={Sze, Vivienne and Chen, Yu-Hsin and Yang, Tien-Ju and Emer, Joel S},
  journal={Proceedings of the IEEE},
  volume={105},
  number={12},
  pages={2295--2329},
  year={2017},
  publisher={Ieee}
}

@article{williams2009roofline,
  title={Roofline: an insightful visual performance model for multicore architectures},
  author={Williams, Samuel and Waterman, Andrew and Patterson, David},
  journal={Communications of the ACM},
  volume={52},
  number={4},
  pages={65--76},
  year={2009},
  publisher={ACM New York, NY, USA}
}

@inproceedings{khalilov2021performance,
  title={Performance analysis of CUDA, OpenACC and OpenMP programming models on TESLA V100 GPU},
  author={Khalilov, Mikhail and Timoveev, Alexey},
  booktitle={Journal of Physics: Conference Series},
  volume={1740},
  number={1},
  pages={012056},
  year={2021},
  organization={IOP Publishing}
}

@article{kitaev2020reformer,
  title={Reformer: The efficient transformer},
  author={Kitaev, Nikita and Kaiser, {\L}ukasz and Levskaya, Anselm},
  journal={arXiv preprint arXiv:2001.04451},
  year={2020}
}

@article{beltagy2020longformer,
  title={Longformer: The long-document transformer},
  author={Beltagy, Iz and Peters, Matthew E and Cohan, Arman},
  journal={arXiv preprint arXiv:2004.05150},
  year={2020}
}

@inproceedings{chen2023large,
  title={Large-Scale Parallelization and Optimization of Lattice QCD on Tianhe New Generation Supercomputer},
  author={Chen, Junlin and Liu, Chaojing and Luana, Zhongzhi and Gong, Ming and Li, Qingfeng and Qian, Depei},
  booktitle={2023 IEEE International Conference on High Performance Computing \& Communications, Data Science \& Systems, Smart City \& Dependability in Sensor, Cloud \& Big Data Systems \& Application (HPCC/DSS/SmartCity/DependSys)},
  pages={499--506},
  year={2023},
  organization={IEEE}
}

@article{guo2025deepseek,
  title={Deepseek-r1: Incentivizing reasoning capability in llms via reinforcement learning},
  author={Guo, Daya and Yang, Dejian and Zhang, Haowei and Song, Junxiao and Zhang, Ruoyu and Xu, Runxin and Zhu, Qihao and Ma, Shirong and Wang, Peiyi and Bi, Xiao and others},
  journal={arXiv preprint arXiv:2501.12948},
  year={2025}
}

@inproceedings{kwon2023efficient,
  title={Efficient memory management for large language model serving with pagedattention},
  author={Kwon, Woosuk and Li, Zhuohan and Zhuang, Siyuan and Sheng, Ying and Zheng, Lianmin and Yu, Cody Hao and Gonzalez, Joseph and Zhang, Hao and Stoica, Ion},
  booktitle={Proceedings of the 29th symposium on operating systems principles},
  pages={611--626},
  year={2023}
}

@misc{TensorRT-LLM,
  author = {{NVIDIA}},
  title  = {{TensorRT-LLM}},
  note   = {\url{https://github.com/NVIDIA/TensorRT-LLM}}
}

@article{walker1996mpi,
  title={MPI: a standard message passing interface},
  author={Walker, David W and Dongarra, Jack J},
  journal={Supercomputer},
  volume={12},
  pages={56--68},
  year={1996},
  publisher={ASFRA BV}
}

@article{koroteev2021bert,
  title={BERT: a review of applications in natural language processing and understanding},
  author={Koroteev, Mikhail V},
  journal={arXiv preprint arXiv:2103.11943},
  year={2021}
}

@article{radford2019language,
  title={Language models are unsupervised multitask learners},
  author={Radford, Alec and Wu, Jeffrey and Child, Rewon and Luan, David and Amodei, Dario and Sutskever, Ilya and others},
  journal={OpenAI blog},
  volume={1},
  number={8},
  pages={9},
  year={2019}
}

@article{brown2020language,
  title={Language models are few-shot learners},
  author={Brown, Tom and Mann, Benjamin and Ryder, Nick and Subbiah, Melanie and Kaplan, Jared D and Dhariwal, Prafulla and Neelakantan, Arvind and Shyam, Pranav and Sastry, Girish and Askell, Amanda and others},
  journal={Advances in neural information processing systems},
  volume={33},
  pages={1877--1901},
  year={2020}
}

@article{liu2024deepseek,
  title={Deepseek-v3 technical report},
  author={Liu, Aixin and Feng, Bei and Xue, Bing and Wang, Bingxuan and Wu, Bochao and Lu, Chengda and Zhao, Chenggang and Deng, Chengqi and Zhang, Chenyu and Ruan, Chong and others},
  journal={arXiv preprint arXiv:2412.19437},
  year={2024}
}

@article{team2024qwen2,
  title={Qwen2 technical report},
  author={Team, Qwen},
  journal={arXiv preprint arXiv:2407.10671},
  year={2024}
}

@article{frantar2022gptq,
  title={Gptq: Accurate post-training quantization for generative pre-trained transformers},
  author={Frantar, Elias and Ashkboos, Saleh and Hoefler, Torsten and Alistarh, Dan},
  journal={arXiv preprint arXiv:2210.17323},
  year={2022}
}

@article{gu2023minillm,
  title={Minillm: Knowledge distillation of large language models},
  author={Gu, Yuxian and Dong, Li and Wei, Furu and Huang, Minlie},
  journal={arXiv preprint arXiv:2306.08543},
  year={2023}
}

@article{ainslie2023gqa,
  title={Gqa: Training generalized multi-query transformer models from multi-head checkpoints},
  author={Ainslie, Joshua and Lee-Thorp, James and De Jong, Michiel and Zemlyanskiy, Yury and Lebr{\'o}n, Federico and Sanghai, Sumit},
  journal={arXiv preprint arXiv:2305.13245},
  year={2023}
}

@article{liu2023scissorhands,
  title={Scissorhands: Exploiting the persistence of importance hypothesis for llm kv cache compression at test time},
  author={Liu, Zichang and Desai, Aditya and Liao, Fangshuo and Wang, Weitao and Xie, Victor and Xu, Zhaozhuo and Kyrillidis, Anastasios and Shrivastava, Anshumali},
  journal={Advances in Neural Information Processing Systems},
  volume={36},
  pages={52342--52364},
  year={2023}
}

@article{xiao2023efficient,
  title={Efficient streaming language models with attention sinks},
  author={Xiao, Guangxuan and Tian, Yuandong and Chen, Beidi and Han, Song and Lewis, Mike},
  journal={arXiv preprint arXiv:2309.17453},
  year={2023}
}

@inproceedings{yu2022orca,
  title={Orca: A distributed serving system for $\{$Transformer-Based$\}$ generative models},
  author={Yu, Gyeong-In and Jeong, Joo Seong and Kim, Geon-Woo and Kim, Soojeong and Chun, Byung-Gon},
  booktitle={16th USENIX Symposium on Operating Systems Design and Implementation (OSDI 22)},
  pages={521--538},
  year={2022}
}

@misc{huggingface_tgi,
  author       = {{Hugging Face}},
  title        = {{Text Generation Inference}},
  howpublished = {\url{https://github.com/huggingface/text-generation-inference}}
}

@inproceedings{sheng2023flexgen,
  title={Flexgen: High-throughput generative inference of large language models with a single gpu},
  author={Sheng, Ying and Zheng, Lianmin and Yuan, Binhang and Li, Zhuohan and Ryabinin, Max and Chen, Beidi and Liang, Percy and R{\'e}, Christopher and Stoica, Ion and Zhang, Ce},
  booktitle={International Conference on Machine Learning},
  pages={31094--31116},
  year={2023},
  organization={PMLR}
}

@inproceedings{zhong2024distserve,
  title={$\{$DistServe$\}$: Disaggregating prefill and decoding for goodput-optimized large language model serving},
  author={Zhong, Yinmin and Liu, Shengyu and Chen, Junda and Hu, Jianbo and Zhu, Yibo and Liu, Xuanzhe and Jin, Xin and Zhang, Hao},
  booktitle={18th USENIX Symposium on Operating Systems Design and Implementation (OSDI 24)},
  pages={193--210},
  year={2024}
}

\begin{IEEEbiography}[{\includegraphics[width=1in,height=1.25in,clip,keepaspectratio]{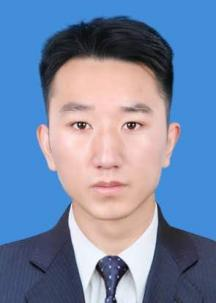}}] {Yao Lu}
 born in 1998, PhD candidate with Beihang University, Beijing China. His main research interests are parallel computing and high-performance computing.
\end{IEEEbiography}
\vspace{-1mm}
\begin{IEEEbiography}[{\includegraphics[width=1in,height=1.25in,clip,keepaspectratio]{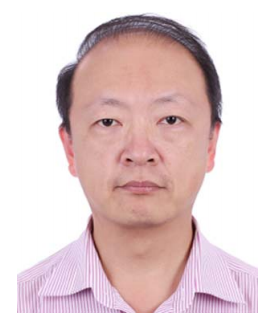}}] {Zhongzhi Luan}
(Member, IEEE) is an Associate Professor with the School of Computer Science and Engineering, Beihang University. He has been engaged in teaching and researching distributed computing, high-performance computing, and computer architecture for many years. His main research interests include resource management in distributed environments, supporting methods and technologies for application development on heterogeneous computing systems, performance analysis, and optimization of high-performance computing applications.
\end{IEEEbiography}
\vspace{-1mm}
\begin{IEEEbiography}[{\includegraphics[width=1in,height=1.25in,clip,keepaspectratio]{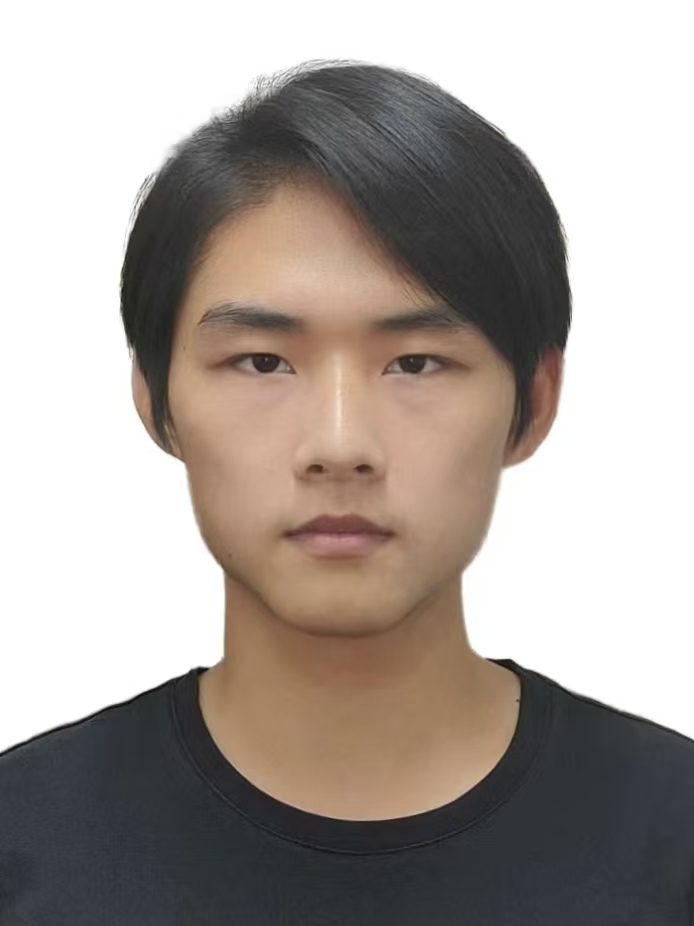}}] {Gen Li}
received the B.S. degree in computer science and technology from the School of Computer Science at Beihang University, Beijing, China, where he is currently pursuing the M.S. degree. His main research interest is high-performance computing.
\end{IEEEbiography}
\vspace{-1mm}
\begin{IEEEbiography}[{\includegraphics[width=1in,height=1.25in,clip,keepaspectratio]{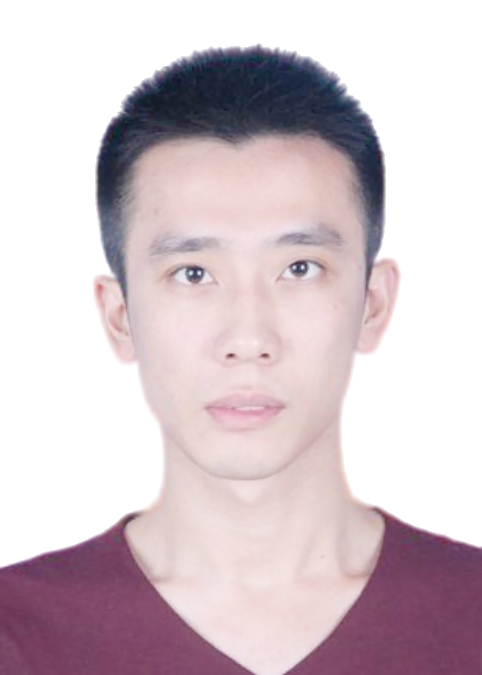}}] {Jiaxing Qi}
received the B.S. degree in software engineering from the Industrial and Commercial College, Hebei University, Baoding, China, in 2017. And the M.S. degree in software engineering from Hebei University, Baoding, China, in 2020. He is currently pursuing the Ph.D. degree at the School of Computer Science and Engineering, Beihang University, Beijing, China. His main research interests include text analytics, and natural language processing.
\end{IEEEbiography}
\vspace{-1mm}
\begin{IEEEbiography}[{\includegraphics[width=1in,height=1.25in,clip,keepaspectratio]{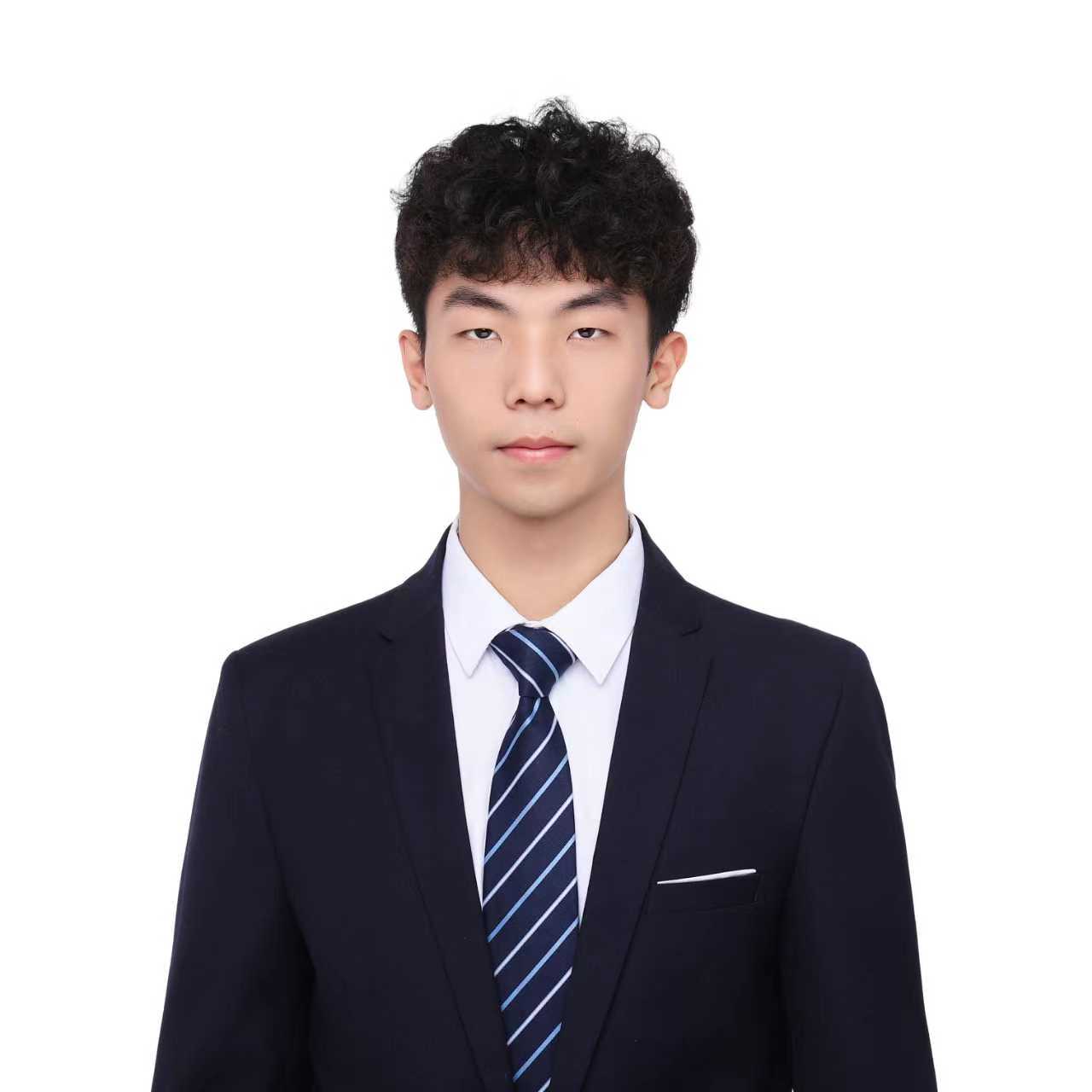}}] {Shiqing Ma}
received the B.S. degree in computer science and technology from the School of Computer Science at Beihang University, Beijing, China, where he is currently pursuing the M.S. degree. His main research interest is high-performance computing.
\end{IEEEbiography}
\vspace{-1mm}
\begin{IEEEbiography}[{\includegraphics[width=1in,height=1.25in,clip,keepaspectratio]{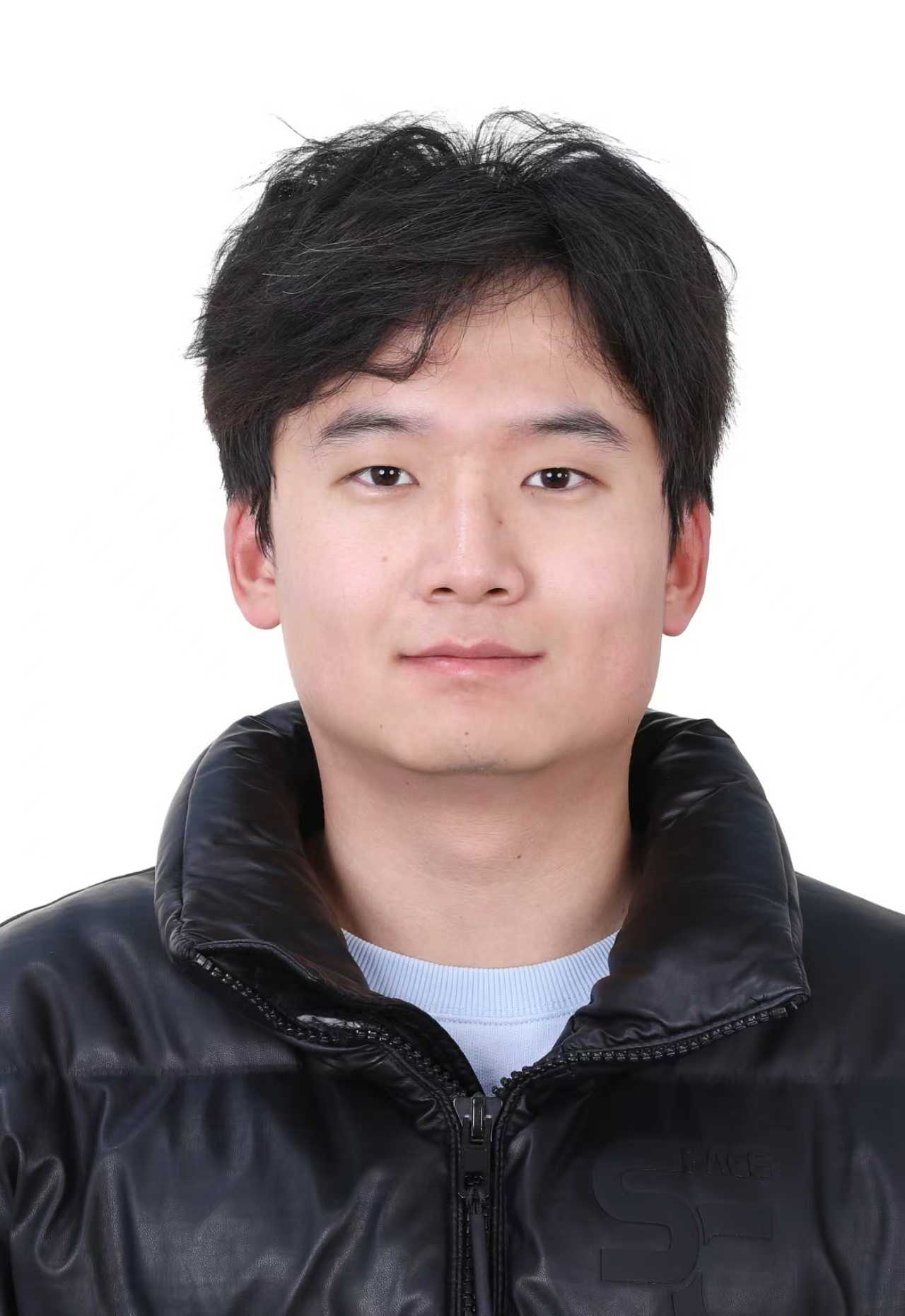}}] {Bin Han}
 received the B.S. degree in computer science from the School of Computer Science at Beihang University, Beijing, China, where he is currently pursuing his PhD degree. His main research interests include high-performance computing and task scheduling systems.
\end{IEEEbiography}
\vspace{-1mm}
\begin{IEEEbiography}[{\includegraphics[width=1in,height=1.25in,clip,keepaspectratio]{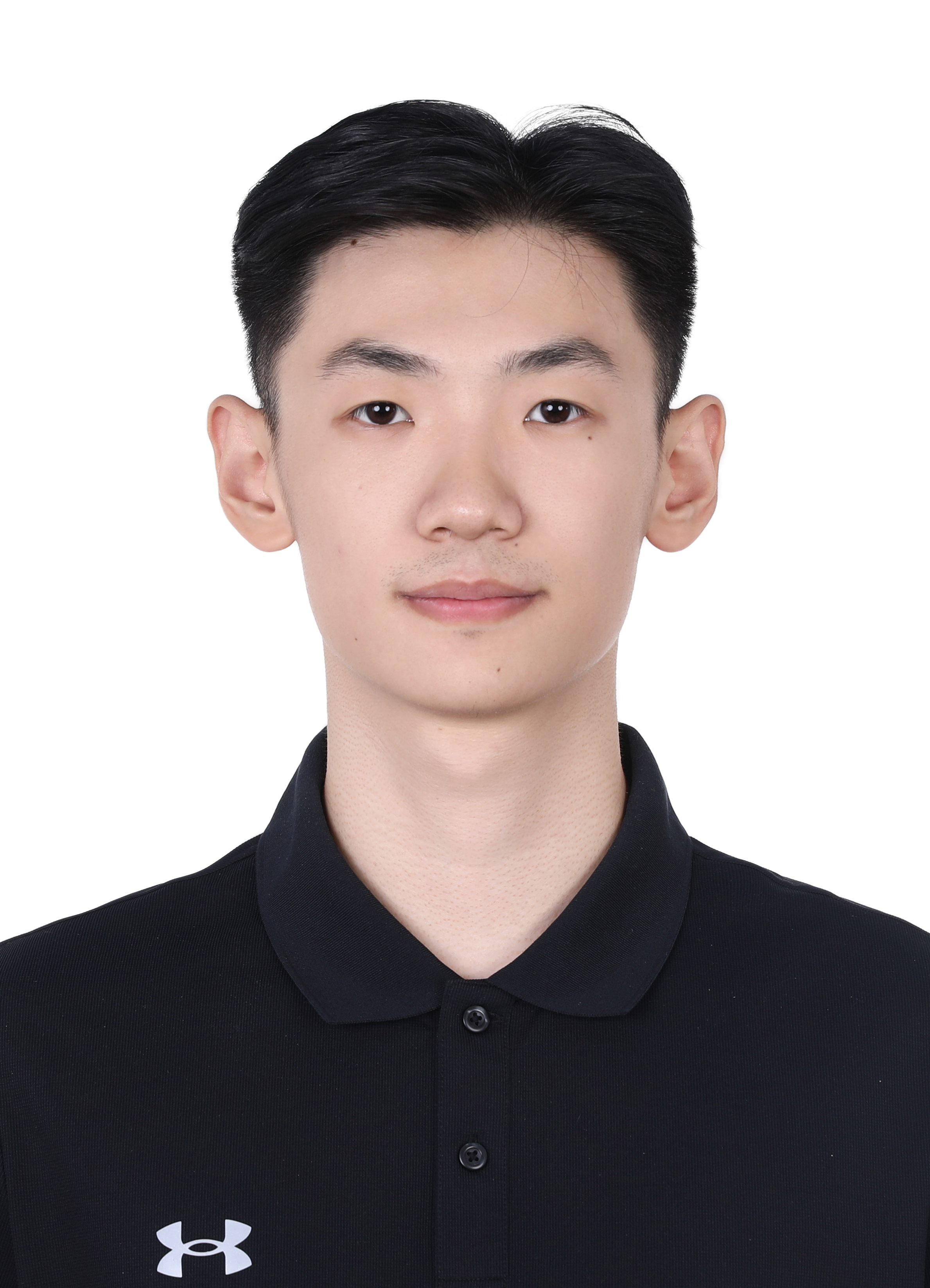}}] {Shizhe Shang}
 received the B.S. degree in information and computing science from Beihang University, Beijing, China, where he is currently pursuing the Ph.D degree. Besides, he attended a double-degree program in Applied Math with Centrale Nantes, Nantes, France. His main research interests include high-performance computing, performance analysis and machine learning.
\end{IEEEbiography}
\vspace{-1mm}
\begin{IEEEbiography}[{\includegraphics[width=1in,height=1.25in,clip,keepaspectratio]{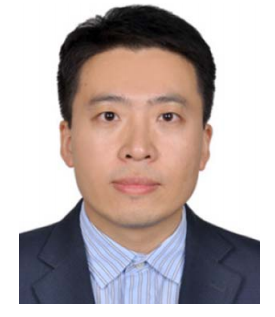}}] {Hailong Yang}
(Member, IEEE) received the Ph.D. degree from the School of Computer Science and Engineering, Beihang University, in 2014, where he is currently working as an Associate Professor. He has been involved in several scientific projects, such as performance analysis for big data systems and performance optimization for large scale applications. His research interests include parallel and distributed computing, HPC, performance optimization, and energy efficiency. He is a Member of China Computer Federation.
\end{IEEEbiography}
\vspace{-1mm}

\begin{IEEEbiography}
[{\includegraphics[width=1in,height=1.25in,clip,keepaspectratio]{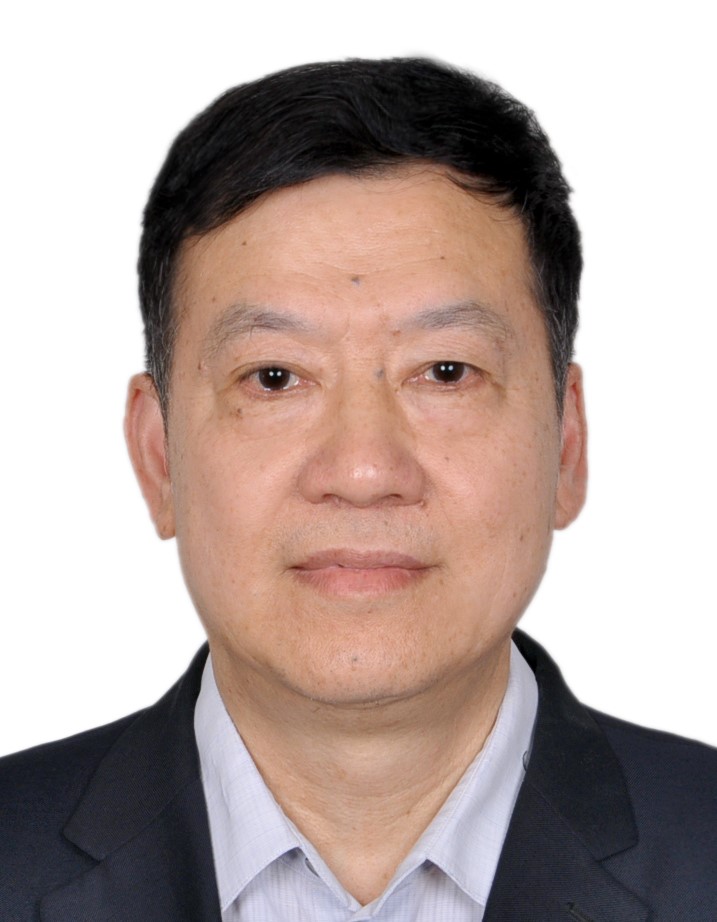}}] {Depei Qian}
received the master's degree from the University of North Texas, Denton, Texas, in 1984. He is a professor with the School of Computer Science and Engineering, Beihang University, China. He served as the chief scientist of China National High Technology Program on high performance computing for 20 years. His research interests include innovative technologies in distributed computing, high performance computing, and computer architecture. He is an academician of Chinese Academy of Science, and also a fellow of the China Computer Federation (CCF).
\end{IEEEbiography}


\vfill

\end{document}